%% file: Rusin.tex
\documentstyle[11pt,aaspp4]{article}
\def\gax{\mathrel{\raise.3ex\hbox{$>$}\mkern-14mu\lower0.6ex\hbox{$\sim$}}}
\def\lax{\mathrel{\raise.3ex\hbox{$<$}\mkern-14mu\lower0.6ex\hbox{$\sim$}}}
\def\gtorder{\mathrel{\raise.3ex\hbox{$>$}\mkern-14mu
             \lower0.6ex\hbox{$\sim$}}}
\def\ltorder{\mathrel{\raise.3ex\hbox{$<$}\mkern-14mu
             \lower0.6ex\hbox{$\sim$}}}

\input psfig

\begin{document}

\title{B1359+154: A Six Image Lens Produced by a z$\simeq$1
Compact Group of Galaxies\footnote{Based on observations made with the
NASA/ESA Hubble Space Telescope, obtained at the Space Telescope Science
Institute, which is operated by AURA, Inc., under NASA contract NAS 5-26555.}}

\author{D. Rusin\altaffilmark{2}, C.S. Kochanek\altaffilmark{3},
M. Norbury\altaffilmark{4}}

\author{E.E. Falco\altaffilmark{3}, C.D. Impey\altaffilmark{5},
J. Leh\'ar\altaffilmark{3}, B.A. McLeod\altaffilmark{3}}

\author{H.-W. Rix\altaffilmark{6}, C.R. Keeton\altaffilmark{5},
J.A. Mu\~noz\altaffilmark{7}, C.Y. Peng\altaffilmark{5}}

\altaffiltext{2}{Department of Physics and Astronomy, University of
Pennsylvania, 209 S. 33rd St., Philadelphia, PA 19104-6396
(email:drusin@hep.upenn.edu)}

\altaffiltext{3}{Harvard-Smithsonian Center for Astrophysics, 60 Garden St.,
Cambridge, MA 02138 (email: ckochanek, efalco, jlehar,
bmcleod@cfa.harvard.edu)}

\altaffiltext{4}{Jodrell Bank Observatory, University of Manchester,
Macclesfield, Cheshire SK11 9DL, UK (email: mn@jb.man.ac.uk)}

\altaffiltext{5}{Steward Observatory, University of Arizona, Tucson, AZ 85721
(email: impey, ckeeton, cyp@as.arizona.edu)}

\altaffiltext{6}{Max-Planck-Institut fuer Astronomie, Koenigsstuhl 17, D-69117
Heidelberg, Germany (email: rix@mpia-hd.mpg.de)}

\altaffiltext{7}{Instituto de Astrofisica de Canarias, Via Lactea, E38200 La
Laguna, Tenerife, Spain (email: jmunoz@ll.iac.es)}

\begin{abstract}

HST V and I-band observations show that the gravitational lens B1359+154
consists of six images of a {\it single} $z_s=3.235$ radio source and its
star-forming host galaxy, produced by a compact group of galaxies at $z_l
\simeq 1$.  VLBA observations at 1.7 GHz strongly support this conclusion,
showing six compact cores with similar low-frequency radio spectra. B1359+154
is the first example of galaxy-scale gravitational lensing in which more than
four images are observed of the same background source. The configuration is
due to the unique lensing mass distribution: three primary lens galaxies lying
on the vertices of a triangle separated by $0\farcs7 \simeq 4 h^{-1}$ kpc,
inside the $1\farcs7$ diameter Einstein ring defined by the radio images. The
gravitational potential has additional extrema within this triangle, creating
a pair of central images that supplement the ``standard'' four-image geometry
of the outer components. Simple mass models consisting of three lens galaxies
constrained by HST and VLBA astrometry naturally reproduce the observed image
positions but must be finely-tuned to fit the flux densities.

\end{abstract}


\section{Introduction}

The vast majority of the $\sim 60$ known arcsecond-scale gravitational lens
systems consist of two or four detectable images, consistent with the generic
lensing properties of smooth, isolated and centrally steep mass distributions
(e.g.\ Blandford \& Kochanek 1987). There are very few cases in which a
``non-standard'' number of images may have been observed.\footnote{B1933+503
(Sykes et al.\ 1998) and B1938+666 (King et al.\ 1997) contain ten and six
lensed radio components, respectively, but this is due to the imaging of a
multi-component source rather than any exotic properties of the lensing
potential.} APM08279+5255 (Ibata et al. 1999) and MG1131+0456 (Chen \& Hewitt
1993) each contain a central component that could be an additional lensed
image created by a sufficiently large galaxy core or shallow mass profile
(Narasimha, Subramanian \& Chitre 1986; Rusin \& Ma 2000). Alternatively,
APM08279+5255 may be a special class of imaging produced by an edge-on disk
(Keeton \& Kochanek 1998), while the central component in MG1131+0456 could be
weak AGN emission associated with the lensing galaxy. In addition, MG2016+112
(Lawrence et al. 1984; Garrett et al. 1996) exhibits a rather complicated
image morphology consisting of four primary components, one of which has three
subcomponents in VLBI maps. This may be a compound lens produced by two
galaxies at different redshifts (Nair \& Garrett 1997).  Benitez et al.\
(1999) suggest an alternative model, but it might not fully explain the VLBI
observations.

Complex interplay between mass distributions can lead to lens systems with
more than four images of a source. For example, ellipsoidal mass distributions
perturbed by shear fields may produce configurations with six or eight images
arranged about the tangential critical curve (Keeton, Mao \& Witt
2000b). However, this requires that the relative magnitudes and orientations
of the internal and external shear axes be finely tuned, and the resulting
cross-sections are quite small. Compound mass distributions have been shown to
be far more efficient at producing a variety of complex geometries in which
five or more images may be formed over a range of radii (Kochanek \&
Apostolakis 1988).  This has been observed in the cluster-lensing case, where
substructure in the gravitational potential created by the resident galaxies
can qualitatively alter the lensing properties that one would expect for a
smooth halo mass distribution (e.g. Natarajan et al. 1998; Meneghetti et
al. 2000). A dramatic example of this is CL0024+1654, which exhibits eight
images of a single blue background galaxy (Kassiola, Kovner \& Fort 1992;
Wallington, Kochanek \& Koo 1995; Colley, Tyson \& Turner 1996; Tyson,
Kochanski \& Dell'Antonio 1998).

Early-type galaxies preferentially participate in lensing (Kochanek et al.\
2000a), so a large fraction of lens galaxies should be members of small groups
and clusters due to the morphology-density relation (Dressler 1980; see also
Keeton, Christlein \& Zabludoff 2000a). Indeed, many lens systems are known to
be perturbed by nearby galaxies (e.g.\ B2319+051; Rusin et al.\ 2000b), groups
(e.g.\ PG1115+080; Schechter et al.\ 1997), or clusters (e.g.\ QSO 0957+561,
Fischer et al.\ 1997; RXJ0921+4529, Mu\~noz et al.\ 2000). A handful of
systems are lensed by more than one primary galaxy (e.g. B1127+385, Koopmans
et al.\ 1999; B1608+656, Koopmans \& Fassnacht 1999), but in these cases
standard image geometries are produced despite merged caustics. Several
additional lenses are observed to have faint satellite galaxies as companions
inside or near the Einstein radius (e.g.\ MG0414+0534, Schechter \& Moore
1993; B1030+074, Xanthopoulos et al.\ 1998; B1152+199, Rusin et al.\ in
preparation). While the likelihood of finding nearly equal mass galaxies close
enough to have merged caustics is predicted to be only $\sim 1\%$ (Kochanek \&
Apostolakis 1988), the probability of finding fainter satellites near the
primary lens is not small because galaxy luminosity functions diverge for
faint galaxies -- all lenses should have faint neighbors, as discussed in the
Appendix. While such systems typically produce regions of the source plane in
which more than four images can form, the small size of the companion galaxies
means that the cross-sections for creating non-standard image geometries are
not significant.

The gravitational lens system B1359+154 (Myers et al.\ 1999), discovered in
the Cosmic Lens All-Sky Survey (CLASS; e.g.\ Myers et al.\ 1995), has been
suspected of containing more than four lensed images. Observations with the
Very Large Array (VLA) and Multi-Element Radio-Linked Interferometer Network
(MERLIN) show a total of six radio components (Myers et al.\ 1999; Rusin et
al.\ 2000a). Four of these components (A--D) are arranged in a typical
quad-lens configuration (maximum image separation of $1\farcs7$), with two
additional components (E and F) residing within the ring defined by the outer
images. Preliminary radio spectral studies of B1359+154 at high frequency
(Myers et al.\ 1999) suggested that E had a slightly flatter spectrum than
A--D, and therefore that the central components may be core-jet emission
associated with a weak AGN in the lensing galaxy or galaxies, as in the case
of B2045+265 (Fassnacht et al.\ 1999). When the spectra were extended down to
5 GHz, however, there appeared to be less disparity among the radio components
(Rusin et al.\ 2000a). Subsequent VLA observations at 15~GHz have failed to
decisively confirm that E is flatter than A--D at high-frequency, or detect
component F. Spectroscopy with the Keck II telescope determined the source
redshift to be $z_s = 3.235$ (Myers et al.\ 1999). Adaptive optics
observations of B1359+154 conducted with the Canada-France Hawaii Telescope
(CFHT) in the infrared K$'$ band (Rusin et al.\ 2000a) detected counterparts
to radio components A--D, and discovered three extended emission peaks
(K1--K3) bracketing the expected positions of E and F. K1--K3 were identified
as three possible lensing galaxies, comprising the core of a compact galaxy
group. Evidence of an arc connecting A, B and C was also observed, along with
a weaker emission feature associated with component E. The compound deflector
system not only explained why attempts to model the outer four components
using a single galaxy had failed (Myers et al.\ 1999), but offered the means
of creating additional extrema in the lensing potential. This opened the
possibility that at least one of the central components is a lensed image.

In this paper we present powerful new evidence from observations with the
Hubble Space Telescope (HST) and Very Long Baseline Array (VLBA) that
B1359+154 consists of six images of a {\em single} background source, lensed
by a compact group of galaxies at $z_l\simeq 1$. In \S2 we present VLBA
observations of B1359+154 and investigate the low-frequency radio spectra of
the components. In \S3 we discuss and analyze HST V and I-band observations,
which offer compelling evidence for the six-image hypothesis. In \S4 we use
preliminary mass modeling to demonstrate that B1359+154 can be naturally
explained as a true six-image lens system. Finally, in \S5 we discuss the
prospects for obtaining improved constraints on the lensing mass distribution
of B1359+154, and ultimately using the system to study the structure of small
galaxy groups at high redshift.

\section{VLBA Imaging of B1359+154}

VLBA observations of B1359+154 were obtained at 1.7 GHz on 1999 December 10
for an on-source integration time of 2 hr, and again on 2000 Aug 28 for 8
hr. These observations were performed by iterating between the target source
for 3 min and the nearby phase-reference calibrator B1413+150 for 1.5 min. The
data for each of the two epochs were calibrated separately within the AIPS
data reduction package, then mapped and model-fit using DIFMAP. The flux
densities of the model-fit components exhibited little variation ($\sim 3\%$)
between epochs, so the two data sets were combined to improve the
signal-to-noise. The resulting naturally-weighted map of B1359+154 is
presented in Fig.\ 1, and has an rms noise level of $45$ $\mu$Jy/beam. All six
radio components seen in previous VLA and MERLIN observations are easily
detected by the VLBA, and maps of the individual components are shown in Fig.\
2. Components A, B, and C each exhibit a compact core with associated jet
emission. Lensing-induced parity reversal is evident in the relative
orientations of the subcomponents. Components D, E and F are each
unresolved. The data were modeled within DIFMAP using a total of 9 gaussian
components (Table 1): 6 compact cores (A1--F1), 1 extended jet (A2) and 2
compact subcomponents (B2, C2).

The compact nature of radio components E and F strongly argue for their
identification as additional lensed images, rather than weak core-jet emission
associated with the lensing mass (Myers et al.\ 1999). First, it would be rare
for jet emission to remain unresolved at $\sim 10$ mas resolution. All but one
pair of radio components shown to be similarly compact in CLASS VLBA follow-up
observations is a pair of lensed images, the only exception being the binary
quasar B0827+525 (Koopmans et al.\ 2000). There is also no evidence in the
VLBA data to suggest that any emission bridge might be connecting E and
F. Second, it is unlikely that both E and F could mark the cores of
independent AGN within the deflector. One would expect the positions of AGN to
be correlated with brightness peaks of the lensing galaxies, and this is not
the case in the CFHT data. Third, the structure of all six radio components
are morphologically consistent with a single background source. Specifically,
if the source consists of a core and weak jet, it is likely that extended
emission would be visible in the more magnified lensed images while the
fainter images show only the compact core.  This is exactly what is observed
in B1359+154: A--C are bright and exhibit weak subcomponents; D--F are faint
and share identical unresolved morphologies.

\begin{figure*}
\psfig{file=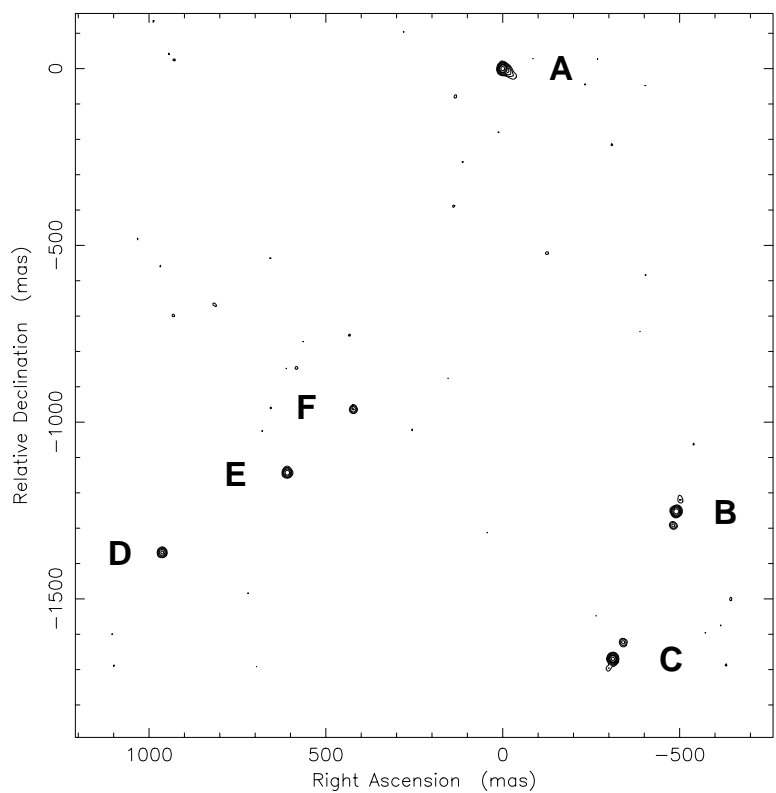,width=4in}
\figurenum{1}
\caption{VLBA 1.7 GHz naturally-weighted map of B1359+154. The data is
restored with a beam of size $13.9 \times 12.5$ mas at PA = $-3.34^{\circ}$. 
The map rms noise level is $45$ $\mu$Jy/beam.}  
\end{figure*}

\begin{figure*}
\begin{tabular}{c c c}
\psfig{file=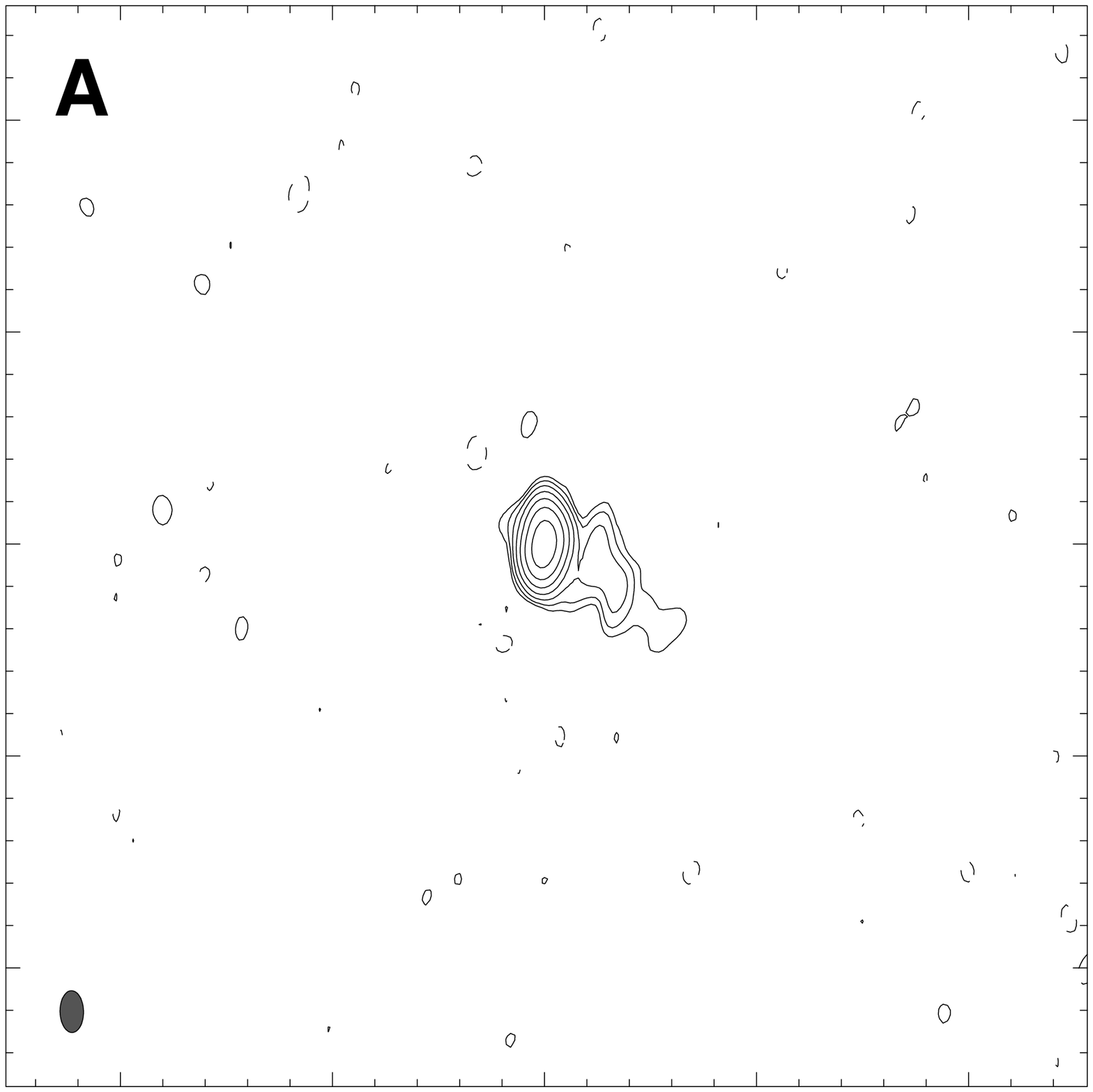,width=2in} & 
\psfig{file=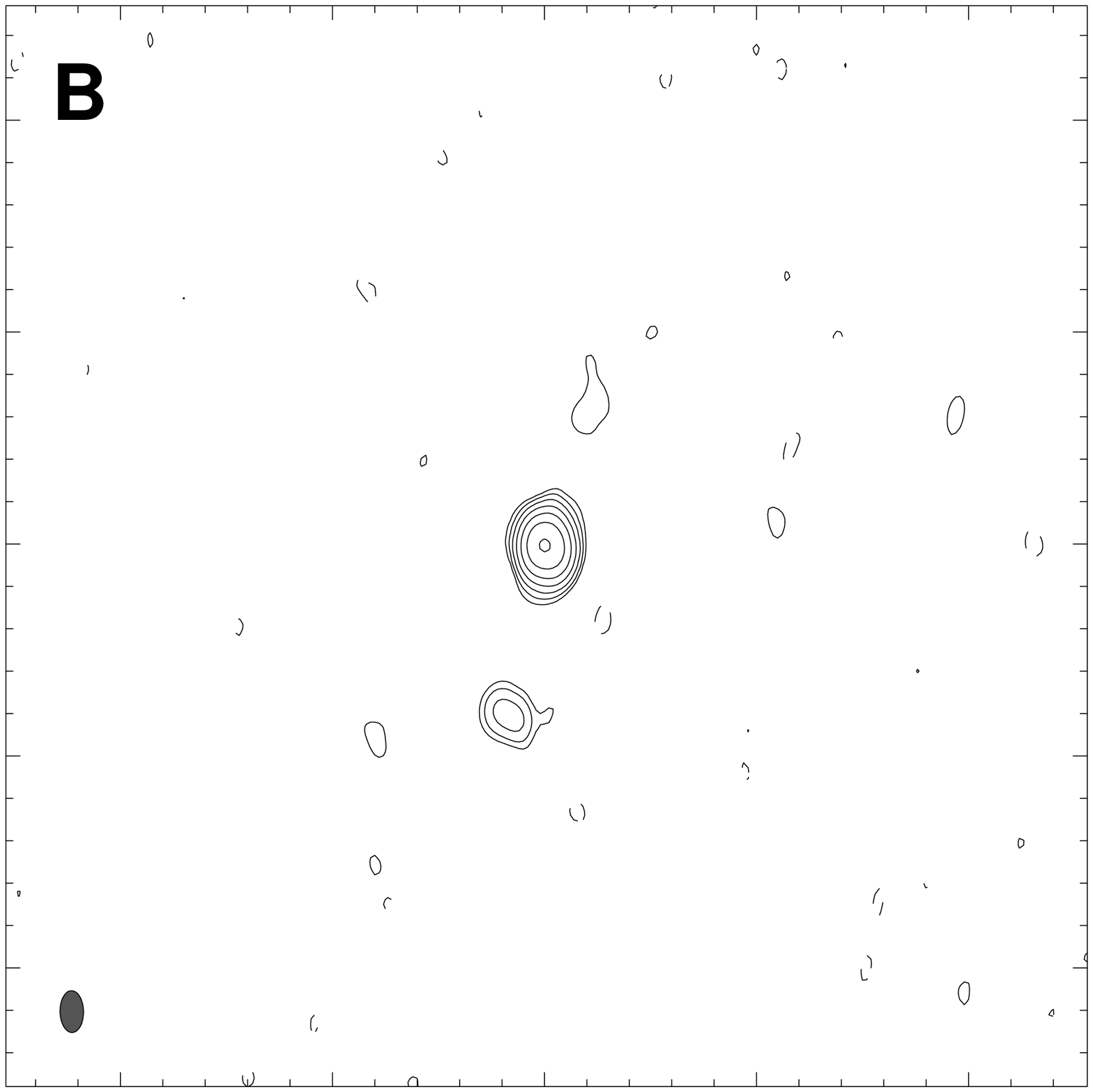,width=2in} & 
\psfig{file=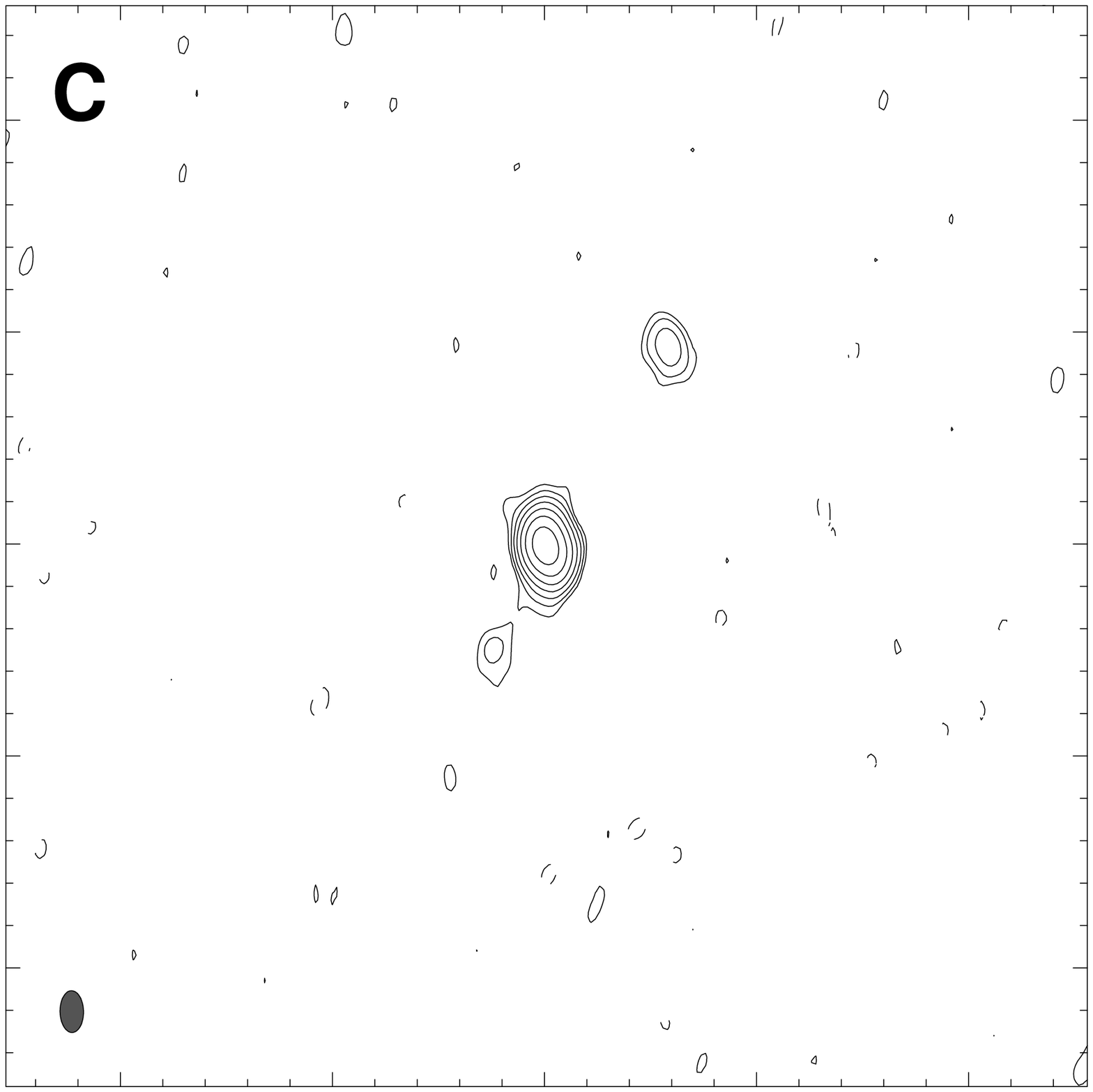,width=2in} \\
\psfig{file=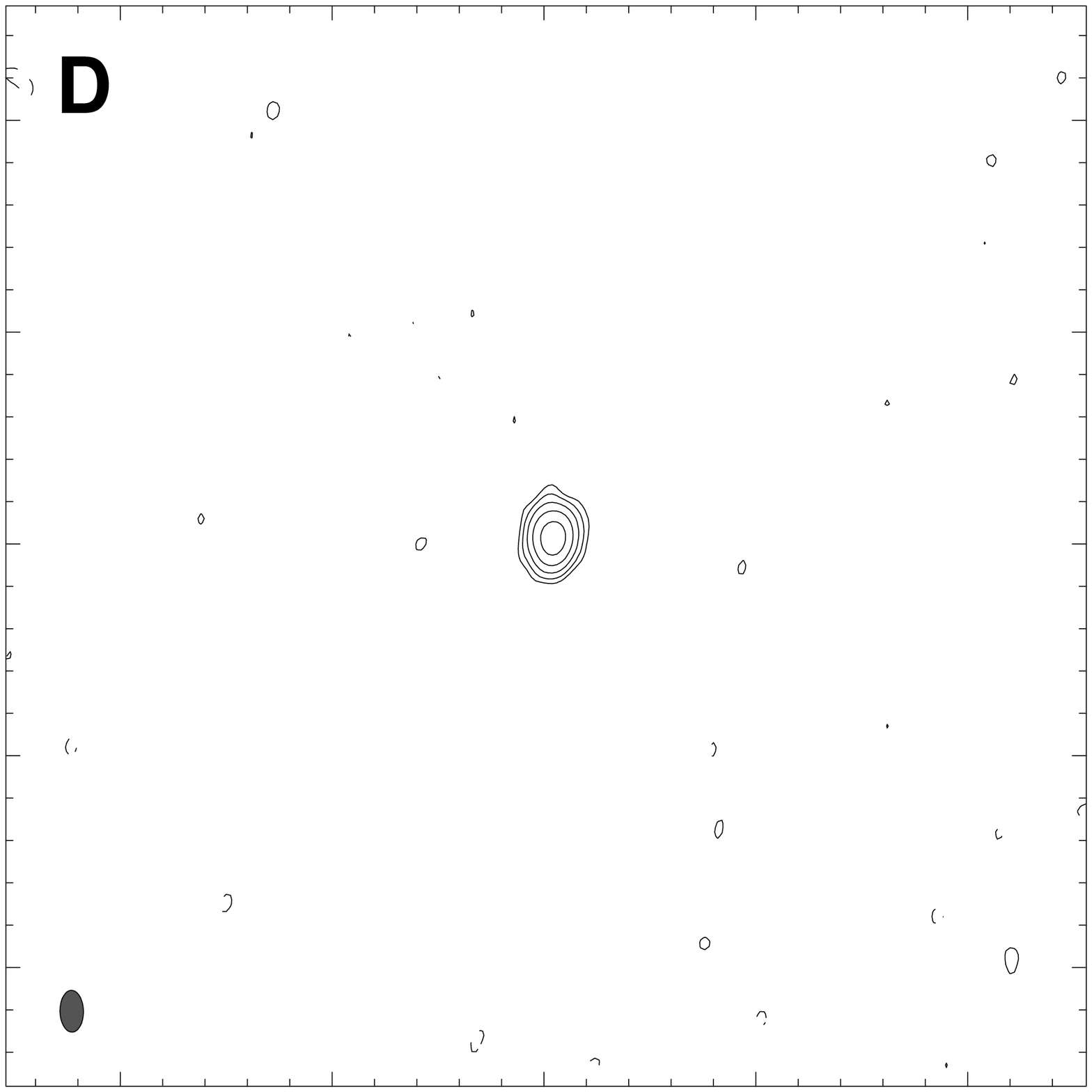,width=2in} & 
\psfig{file=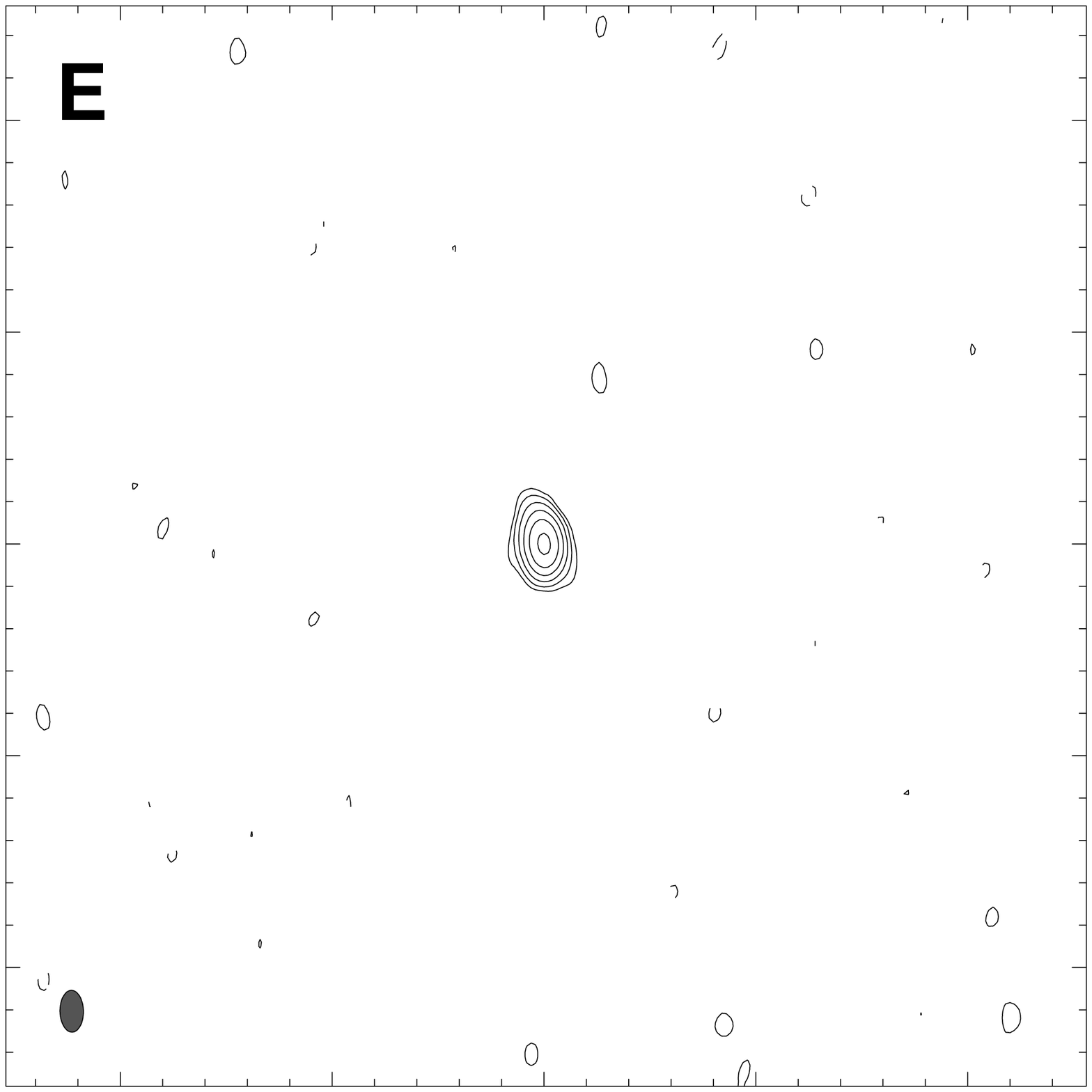,width=2in} & 
\psfig{file=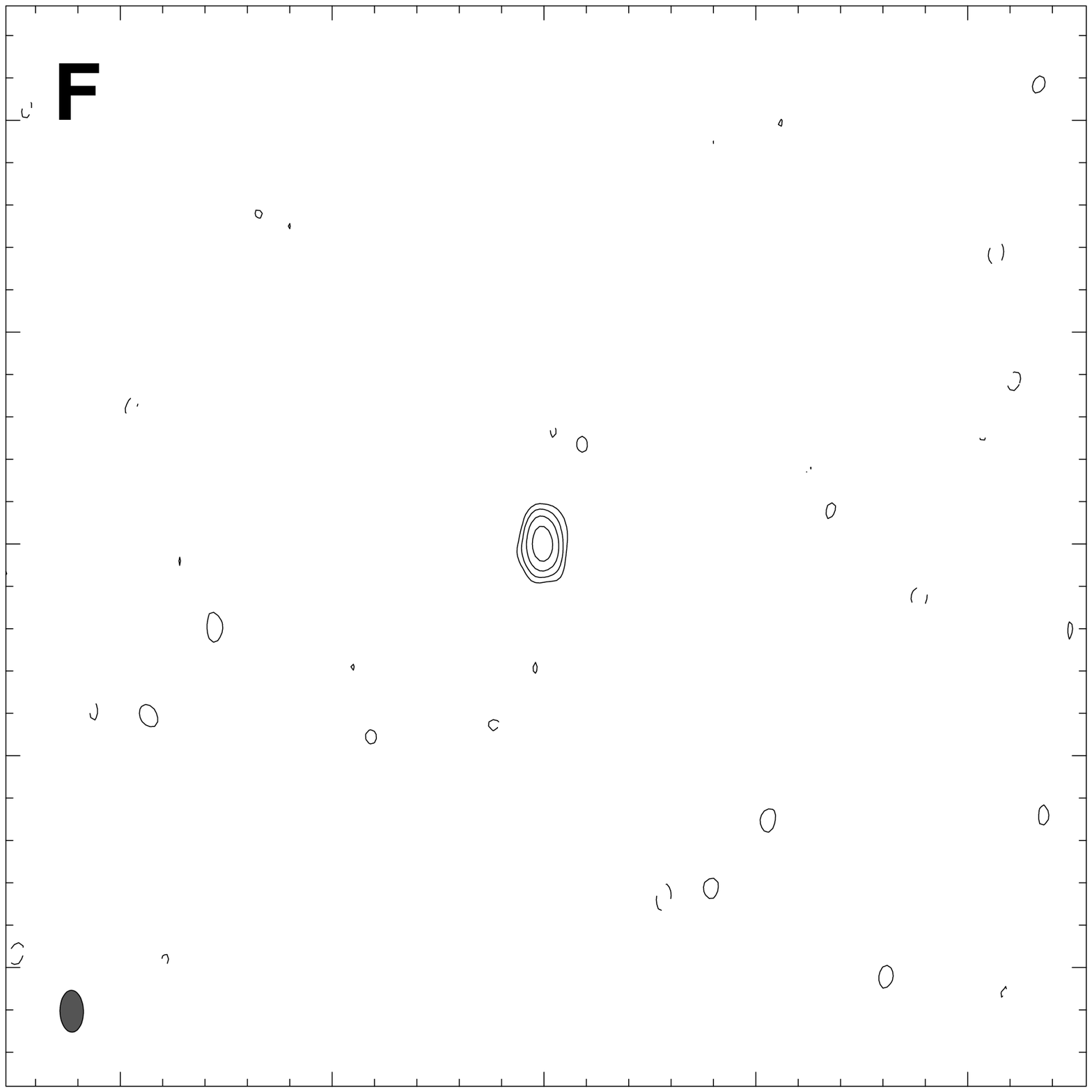,width=2in} \\ 
\end{tabular}
\figurenum{2}
\caption{VLBA 1.7 GHz naturally-weighted maps of B1359+154 components. Top: A,
B, C. Bottom: D, E, F. The beam is $9.88 \times 5.54$ mas at PA =
$1.11^{\circ}$. Each box size is $250 \times 250$ mas. Contours are at $3, 6,
9... \times$ the map rms.}

\end{figure*}

\begin{figure*}
\psfig{file=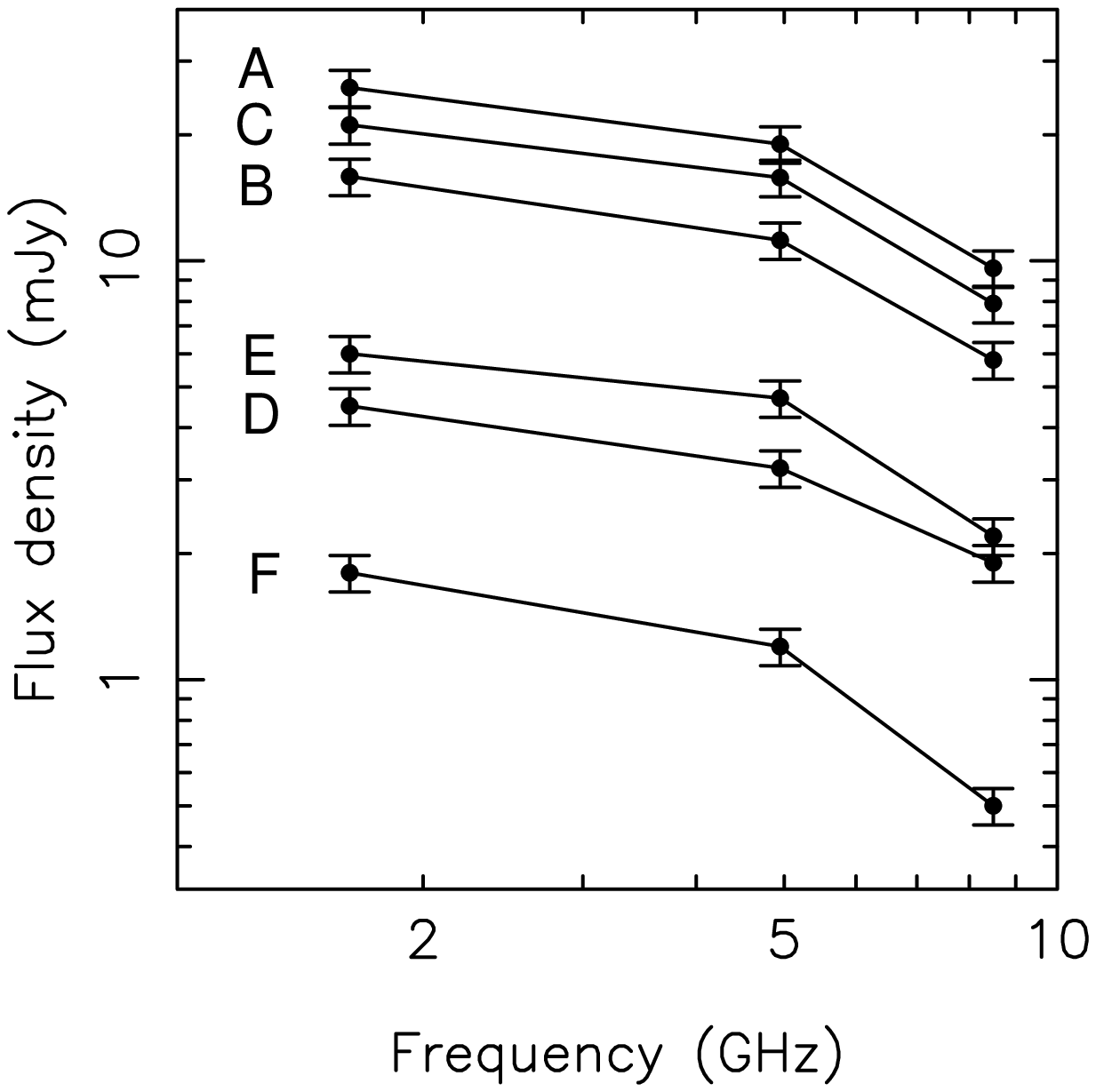,width=4in}
\figurenum{3}
\caption{Radio spectra of B1359+154 components. Data from the 1.7 GHz VLBA, 5
GHz MERLIN (Rusin et al.\ 2000a) and 8.5 GHz VLA (Myers et al.\ 1999)
observations. We assume flux density uncertainties of 5\% to account for
source variability and model-fitting errors.}

\end{figure*}

Next we construct low-frequency radio spectra of the components by combining
the 1.7 GHz VLBA flux densities with those previously obtained at 5 GHz using
MERLIN (Rusin et al.\ 2000a) and at 8.5 GHz using the VLA (Myers et al.\
1999). One need not be concerned about combining data from different epochs as
there is little evidence for any significant variability in the lensed source.
The resulting radio spectra are plotted in Fig.\ 3 and exhibit striking
similarities. Each of the six components has a spectrum that is rather flat
($\alpha_{1.7}^{5} \simeq -0.3$, where $S_{\nu} \propto \nu^{\alpha}$) between
1.7 and 5 GHz, then falls off sharply ($\alpha_{8.5}^{15} \simeq -1.2$)
between 5 and 8.5 GHz. The 1.7 -- 5 GHz spectral indices for all six
components (Table 1) agree within measurement errors. Even if the sharpness of
the 5 GHz spectral break were somehow exacerbated by variability, it would
still require that all six components vary in the same manner -- something
that would be impossible unless E and F are related to A--D. Therefore, the
morphologies and low-frequency spectra of the B1359+154 radio components
strongly suggest that they are six images of a single background source.

\section{HST Imaging of B1359+154}

Using the WFPC2 instrument on HST we obtained images of B1359+154 in the F555W
(``V'', 5200\,sec) and F814W (``I'', 5000\,sec) bands on 2000 July 10.  Each
observation was divided into four subexposures, at two dither positions, which
were used to remove cosmic rays and bad pixels.  Fig.~4 shows the resulting
images.  Optical counterparts to all six radio components are observed, along
with extended emission from the three galaxies. The images were modeled as a
combination of point sources and de Vaucouleurs models following the modeling
procedures of Leh\'ar et al. (2000) for CASTLES observations of other lenses.
Astrometric and photometric results are presented in Table 2. The positions of
images A--F derived from the HST data match the radio positions to an accuracy
of 0\farcs02, consistent with our internal error estimates.

\begin{figure*}
\begin{tabular}{c}
\psfig{file=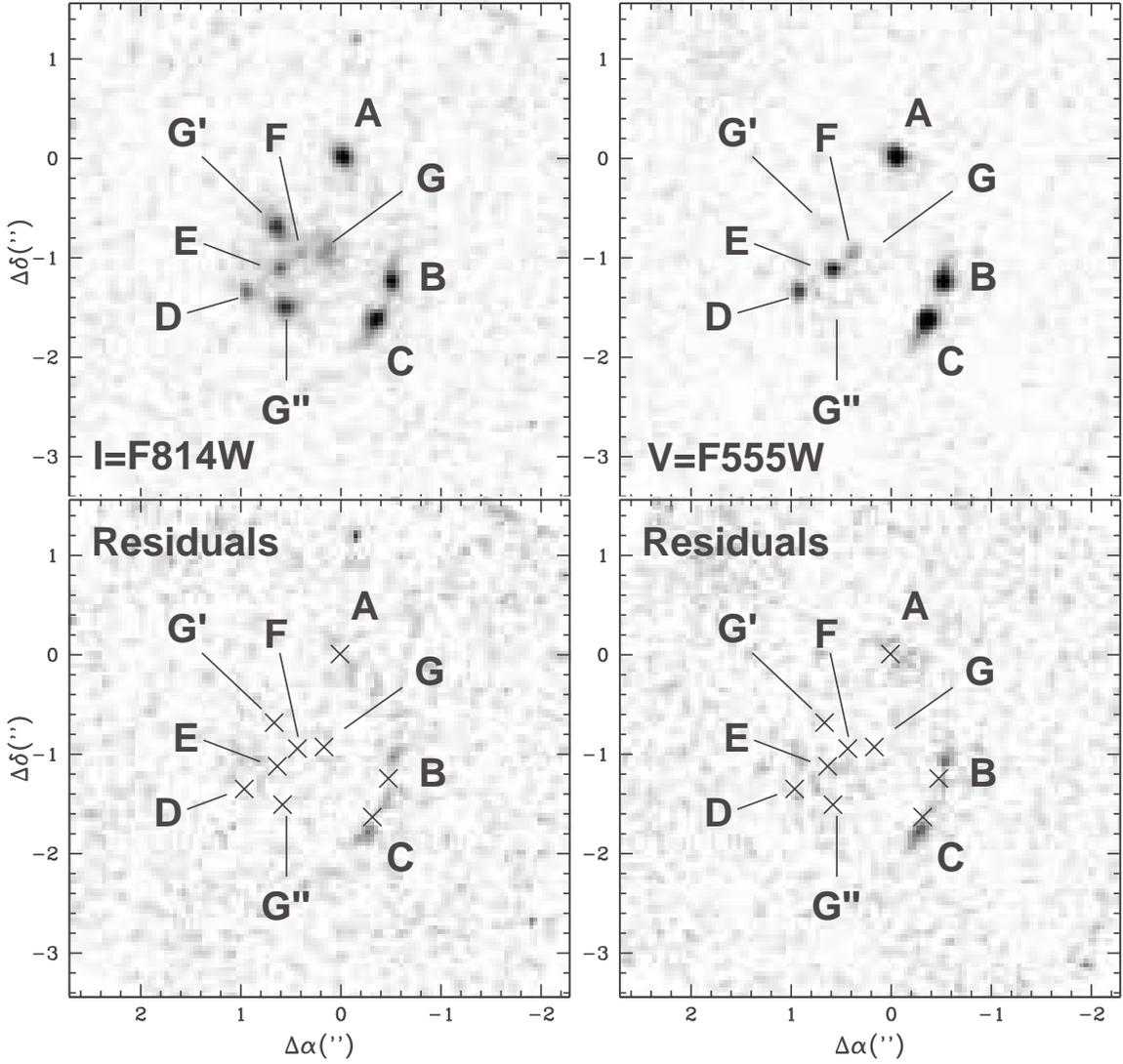,width=6in}\\
\end{tabular}
\figurenum{4}
\caption{HST I-band (left) and V-band (right) images of B1359+154.  The top
panels show the final images with the components labeled as in Table~2.  The
bottom panels show the residuals after subtracting counterparts to the six
radio cores and the three lens galaxies.  Crosses mark the centers of the
subtracted components.  }
\end{figure*}

All six optical counterparts of the radio components are blue, with an average
color of V--I$=0.5$~mag and a dispersion in the colors of only $0.1$~mag.  The
flux ratios of the 6 radio components and their optical counterparts differ by
less than a factor of two between 1.7~GHz and the V-band, so they must be six
lensed images.  Three of the counterparts (A, B and C) are clearly extended in
both the V and I-band images, consistent with the presence of an Einstein ring
in the Rusin et al. (2000a) infrared image.  The rest frame wavelengths of the
V and I-band filters at the source redshift of $z_s=3.235$ are 1300\AA$\,$ and
1900\AA$\,$, respectively, so the host galaxy of the radio source must have a
strong ultraviolet continuum to produce the arcs. While we lack images in
bluer filters, and the Myers et al. (1999) spectrum of the source is
contaminated by AGN activity, the host galaxy appears to be a star-forming
Ly-break galaxy.  The emission lines from the source are relatively narrow
(Ly$\alpha$ and N{\sc V} are well resolved) and lines characteristic of both
AGN ({\sc Civ}) and star formation ({\sc Ciii}$\rbrack$) are seen in the
spectrum.

The optical counterparts (G, G$'$, G$''$) of the three infrared sources K1--K3
are extended and red. We have relabeled the sources using the standard CASTLES
convention in which G is the most luminous lens and G$'$/G$''$ are of lower
luminosity. All three are seen in the I-band image and essentially vanish in
the V-band image, with a color of V--I$\gtorder 3$~mag. They are clearly three
distinct galaxies which are well modeled by de Vaucouleurs profiles, rather
than compact star-forming regions in a larger galaxy.  With separations of
only $4h^{-1}$~kpc (for $z_l \simeq 1.0$ in an $\Omega_0=0.3$ flat
cosmological model), they are likely to be the core of a compact group.
Unlike other lenses with secondary lens galaxies, the three galaxies have
similar luminosities.  Redshift estimates based on their optical colors and
the mass estimates derived from our models of the system (\S4) and the
fundamental plane (Kochanek et al.\ 2000a) are mutually consistent.  We find
$z_l=1.35\pm0.16$, $0.88\pm0.06$ and $0.94\pm0.07$ for galaxies G, G$'$, and
G$''$, respectively.  The average redshift is $z_l=1.05\pm0.20$ and the
weighted average is $z_l = 0.94 \pm 0.04$, but for simplicity we will just
assume $z_l=1$ for future estimates.  The scatter of the individual estimates
is consistent with the scatter found by Kochanek et al. (2000a) for lenses
with known redshifts.  Such a group of galaxies is unlikely to be a projection
effect given the separations and the size of the overall lens sample. Based on
the results in the Appendix we estimate that the {\it a posteriori}
probability of finding such a chance projection in a sample of 60 lenses is
$\sim 10^{-3}$.  We conclude that the lens is the core of a compact galaxy
group (see Hickson 1997 for a review of compact groups).  With three lens
galaxies of comparable luminosity and redshift inside their joint Einstein
ring, the gravitational potential should have additional extrema inside the
triangle formed by the galaxies.  While images A--D resemble a standard
four-image lens morphology, two additional images (E and F) lie in the region
between the three primary lenses.  {\it Therefore, HST observations offer
powerful evidence that B1359+154 is the first example of a true six-image
galaxy-scale gravitational lens system.}

We also cataloged the nearby galaxies in the HST images using the SExtractor
package (Bertin \& Arnouts 1996).  The software classified the optical
objects, and computed total magnitudes using Kron-type automatic apertures
(see Leh\'ar et al. 2000).  Colors were computed using fixed apertures scaled
to the F814W size of each object.  Table~3 lists all the galaxies detected by
HST within 20\arcsec\ of the lens, after visually confirming the SExtractor
classification.  The galaxies were labeled G\# in order of decreasing I-band
brightness, or K\# for the galaxies found in the Rusin et al. (2000a) infrared
observations which were not detected in the I-band image.  Table~3 also
includes an estimate of the shear perturbation each neighboring galaxy can
produce on the lens assuming they lie at the same redshift and have the same
mass-to-light ratios (see Leh\'ar et al. 2000).  The sum of the shear
contributions is $\gamma=0.074$ at $\theta_{\gamma}=83^\circ$, although the
largest source of shear, galaxy G1, is probably a foreground galaxy.  The
contribution of all the other galaxies is $\gamma=0.049$ at
$\theta_{\gamma}=-60^\circ$.

\section{Preliminary Mass Models}

Constructing any believable model for the lensing mass of B1359+154 is a
significant challenge. At a minimum the model must include mass distributions
to represent each of the three lensing galaxies plus environmental
contributions to the lensing potential, either in the form of external shear
or a separate group halo. Since even the simplest models require a large
number of free parameters, the best hope for producing a realistic mass model
may ultimately rely on constraints from deep HST imaging of the extended arc
emission (Kochanek, Keeton \& McLeod 2000b). For now our primary objective is
to demonstrate that a six-image lens system can be naturally produced by the
observed lensing galaxies.

All models are constrained with 12 coordinates and 5 flux density ratios of
the lensed radio components, as derived from the VLBA data listed in
Table~1. Only the core components A1--F1 are used at this time. We set a fit
tolerance of 1 mas for the image positions and assume a 10\% uncertainty on
the flux densities. The galaxy positions are constrained according to the
astrometric error bars given in Table 2. The goodness-of-fit parameter is
therefore $\chi^2_{tot}=\chi^2_{pos}+\chi^2_{flx}+\chi^2_{gal}$, which
includes contributions to the fit from the positions of the radio components,
($\chi^2_{pos}$), their flux ($\chi^2_{flx}$), and the positions of the lens
galaxies ($\chi^2_{gal}$). The redshift of the group is set at $z_l = 1$, and
a flat cosmology with $\Omega_o = 0.3$ is assumed for all calculations. The
models are optimized using the {\it lensmodel} software package (Keeton
2000).\footnote{The code is publicly available at
http://penedes.as.arizona.edu/~ckeeton/gravlens}

\subsection{A Demonstration of Principle}

Smooth gravitational potentials must produce an odd number of lensed images
(Burke 1981). However, images can be trapped and demagnified in the singular
or near-singular cores of lensing galaxies (Narasimha et al.\ 1986). We can
label the possible image configurations by A/B, where A is the true number of
images and B is the observed number of images. Since A is always odd,
B1359+154 must be either a 7/6 or 9/6 system.  A single lens can generate a
7/6 image configuration, but it should consist of 6 images near the Einstein
radius of the system (see Keeton et al. 2000b). The central location of images
E and F and the comparable luminosities of the three galaxies (1.0:0.4:0.4 for
G:G$'$:G$''$) strongly suggest that all three lenses are important and that we
are looking at a 9/6 lens system.

We begin by considering a compound mass distribution consisting of three
singular isothermal spheres (SIS) in an external shear field. This model is
unlikely to fit the data very well as it ignores galaxy ellipticity, but we
use it to demonstrate the principles of compound deflectors in the context of
B1359+154. From the expected caustic structure alone one can argue that 9/6
systems are possible. Each SIS deflector has an associated radial caustic on
the source plane, at which an inward-crossing source produces a pair of images
at the corresponding galaxy center: one observable and one trapped in the
core. The asymmetry of the potential will also lead to at least one tangential
caustic, at which an inward-crossing source produces a pair of observable
images at the corresponding critical curve. Therefore, a 9/6 system is formed
when a source sits within one tangential and three radial caustics. Because
three positive parity images will be trapped in the galaxy cores, four of the
observed images must have negative parity (saddle points in the time delay
surface) to ensure that the parities of all nine images sum to 1 (Burke 1981;
Blandford \& Narayan 1986). Images A--D are arranged in a typical quad
configuration, so we expect their parities to alternate in the usual way:
A(+), B(--), C(+), D(--). This means that both E and F must correspond to
saddle points (--), and we model the system under this assumption.

The critical curves, caustics and time delay surface for the best-fit 3SIS
model are plotted in Fig.\ 5. Because the separation of the galaxies is small
compared to their Einstein radii, a single tangential critical curve encloses
all three galaxies (Fig.\ 5a). This maps to the highly distorted tangential
(astroid) caustic on the source plane (Fig.\ 5b). The large circular features
in Fig.\ 5b correspond to the radial caustics associated with each SIS. These
not only significantly overlap each other, but also overlap much of the
astroid caustic creating a substantial 9/6 region in which the source
resides. The cross-section of the model is dominated by 3/2, 5/3, 7/5 and 9/6
geometries, but it can also produce 7/4 and 11/8 configurations. The 11/8
region is due to the tangential caustic folding over on itself and creating a
swallowtail catastrophe. The probability of observing 7/4, 7/5, 9/6 and 11/8
systems will be greatly enhanced through magnification bias.  Despite the
complexity of the caustic structure, the topology of the virtual time delay
surface (Fig.\ 5c) is quite generic (Schneider 1985; Blandford \& Narayan
1986). Images A--D have the standard topology for four-image lenses.  Images
A--C are associated with a {\it lemniscate} critical contour encompassing two
minima (A and C) and the saddle point (B).  Image D is a saddle point
associated with a {\it lima\c con} contour which would usually encompass
images A--C in the large loop and the core of the lens galaxy in the smaller
loop.  In this case, however, the smaller loop contains all three lens
galaxies.  Images E and F lie at the saddle points of two new {\it lemniscate}
critical contours with maximum-saddle-maximum topologies, where the maxima sit
on the galaxy cores.

The 3SIS model has a best-fit of $\chi^2 = 1570$ for NDF = 10, and its
parameters are listed in Table 4. The model poorly fits the positions of
images A ($\chi^2_{pos,A} = 323$), B ($\chi^2_{pos,B} = 659$) and F
($\chi^2_{pos,F} = 299$). The flux density of image A is also significantly
lower than observed (predicted 2.4 mJy, $\chi^2_{flx,A} = 80$). Given the
simplicity of the model, the errors in the image positions are of little
concern and should be greatly improved by adding ellipticity or adjusting the
radial mass profiles of the galaxies. The mismatch of the flux densities is a
more generic problem. Components B and C are clearly a merging pair of images
associated with the tangential critical curve, and in most lens systems would
be the brightest images. To reproduce the observed flux ordering in B1359+154,
the critical curve must pass closer to A than either B or C, so that A will be
the most magnified.\footnote{It is conceivable that the brightness of A may be
produced by microlensing due to substructure in the lensing mass distribution
(Schneider \& Mao 1998). However, the large required amplification makes this
less likely.} The 3SIS model is unable to accomplish this, as A sits quite far
from the tangential critical curve (Fig.\ 5a). Accounting for the relative
fluxes will therefore require models with more free parameters.

\begin{figure*}
\begin{tabular}{c c}
\psfig{file=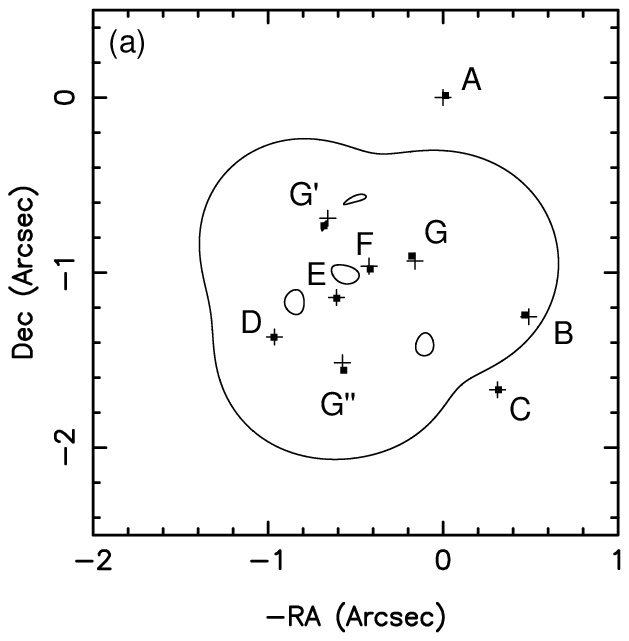,width=3.3in} & 
\psfig{file=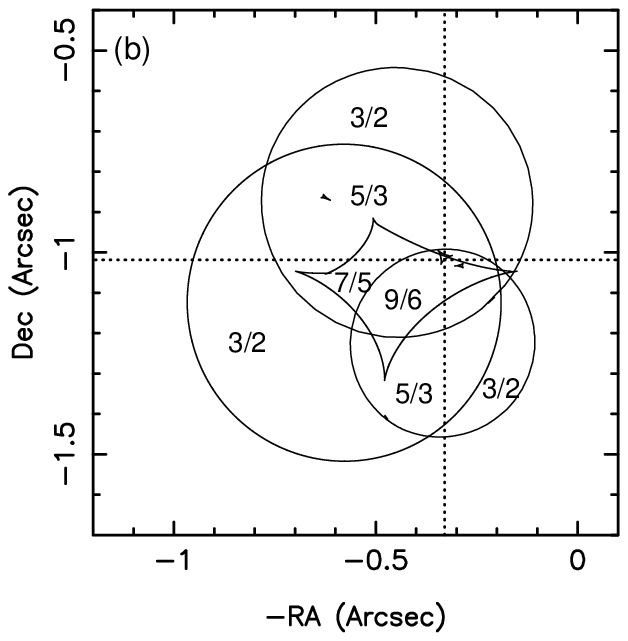,width=3.3in}\\ 
\psfig{file=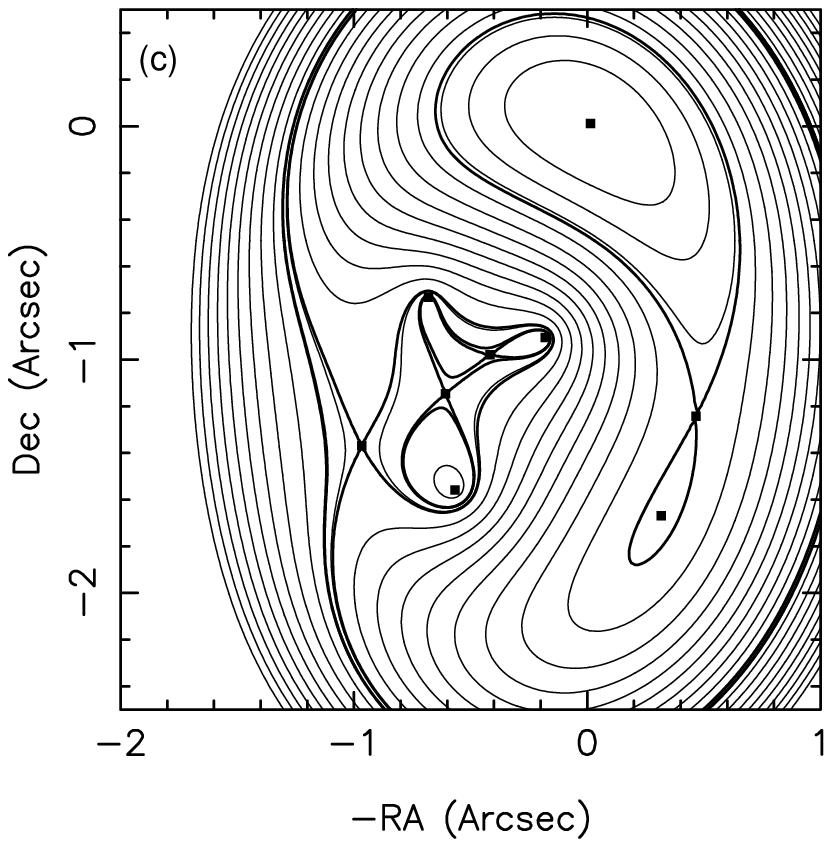,width=3.3in} & \\ 
\end{tabular}
\figurenum{5}
\caption{Best-fit 3SIS model. (a) Critical curves. Observed (model-predicted)
image and galaxy positions are marked as crosses (squares). Note that A is far
from the tangential critical curve, and thus cannot be highly magnified in
this model. (b) Caustic curves. Prominent imaging regions are marked as A/B,
where A is the total number of images and B is the observable number of
images. The source sits at the intersection of the dotted lines, near a
swallowtail catastrophe. (c) Contours of constant time delay, in steps of $3
h^{-1}$ days outward from A (thin lines), except for the critical contours
(bold lines) passing through the saddle points in this surface (B, D, E and
F). Images trapped in the galaxy cores correspond to local maxima.}

\end{figure*}

\subsection{More Realistic Models}

We now attempt to improve the fits to B1359+154 by investigating more
realistic lens models. The mass distributions of lensing galaxies are
ellipsoidal rather than spherical, so we consider models with 3 singular
isothermal ellipsoids (SIE) in an external shear field. The SIE is described
by the scaled surface mass density
\begin{eqnarray}
\kappa(x, y) = {b/2 \over [(1-\epsilon) x^2 + (1+\epsilon)y^2]^{1/2}}
\end{eqnarray}
where $b$ is the critical radius and $\epsilon = (1-f^2)/(1+f^2)$ is the
ellipticity, with $f$ as the mass axial ratio. In one set of trials we fix the
orientations of G, G$'$ and G$''$ according to the de Vaucouleurs surface
brightness profiles listed in Table 2. We refer to this as the 3SIE/FIX
model. However, because these orientations are not very well constrained by
HST, we also run trials in which the position angles of the galaxies are left
as free parameters (3SIE model). Finally, we recognize that galaxies forming
the core of a compact group may be truncated on the scale of the inter-galaxy
separation. We thus consider pseudo-Jaffe (PJ) models (Keeton \& Kochanek
1998), which have a truncation radius at which the density breaks from $\rho
\propto 1/r^2$ to $1/r^4$. The PJ profile is described by
\begin{eqnarray}
\kappa(x, y) = {b/2 \over [(1-\epsilon) x^2 + (1+\epsilon)
y^2]^{1/2}} - {b/2 \over [(1-\epsilon) x^2 + (1+\epsilon)
y^2 + a^2]^{1/2}}
\end{eqnarray} 
We set $a = 0\farcs3$, roughly half the inter-galaxy separation in this
system, and investigate models with spherical (3PJS) and elliptical (3PJE/FIX,
3PJE) pseudo-Jaffe deflectors.

The 3SIE/FIX model has a fit statistic of $\chi^2_{tot} = 216$ for NDF = 7,
significantly improving upon the 3SIS model with regard to the image positions
($\chi^2_{pos} = 4$). The flux densities are still poorly matched to their
observed values ($\chi^2_{flx} = 182$).  The model-predicted flux densities of
A--C are each much fainter than observed, and the fit to image A is
particularly poor (predicted 2.8 mJy, $\chi^2_{flx,A} = 77$).  It is unlikely
that any fine-tuning of the model will remove this discrepancy so long as the
galaxy position angles remain fixed.

The 3SIE model relaxes constraints on the galaxy orientations and provides the
best fit of any model we investigate, $\chi^2_{tot} = 43$ for NDF = 4. The
critical curves, caustics and time delay surface are plotted in Fig.\ 6. In
this model the critical curve passes closer to A than either B or C (Fig.\
6a), ensuring that A will be the brightest image. The key ingredient here is
the placement of the unlensed source, which lies very close to the cusp of a
swallowtail catastrophe inside the tangential caustic (Fig.\ 6b). As the
source approaches this cusp, A diverges in magnification and would split into
three adjacent images upon crossing into the 11/8 region. Therefore A can be
highly magnified by placing a source just outside the swallowtail cusp. The
largest contribution to the 3SIE fit statistic comes from the flux density of
C (predicted 10.6 mJy, $\chi^2_{flx,C} = 22$). This is not a serious concern,
as a slight shifting of the critical curve from B toward C could reproduce the
proper flux ordering, and may be accomplished through unmodelable substructure
in the lensing mass distribution. The second largest contribution is from the
RA of galaxy G ($\chi^2_{gal,G} = 15.6$), which accounts for nearly all of
$\chi^2_{gal} = 16.0$. This is exacerbated by the small estimated uncertainty
($0\farcs008$) of this coordinate. If we have underestimated the error, or had
used circular error bars of radius equal to the major axis of the error
ellipse ($0\farcs024$ for G), $\chi^2_{gal}$ would be much smaller ($\simeq
2$).

\begin{figure*}
\begin{tabular}{c c}
\psfig{file=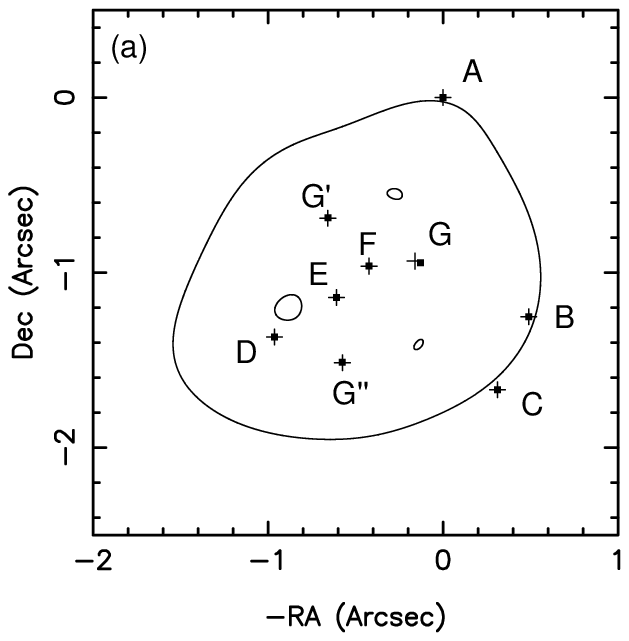,width=3.3in} & 
\psfig{file=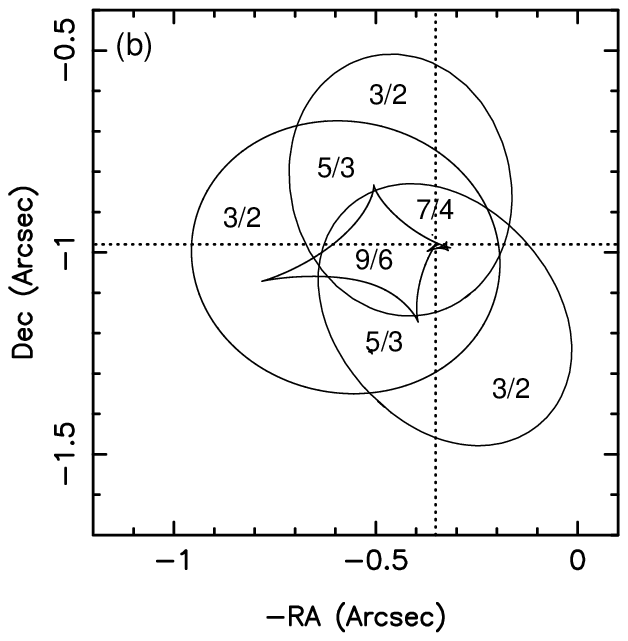,width=3.3in}\\ 
\psfig{file=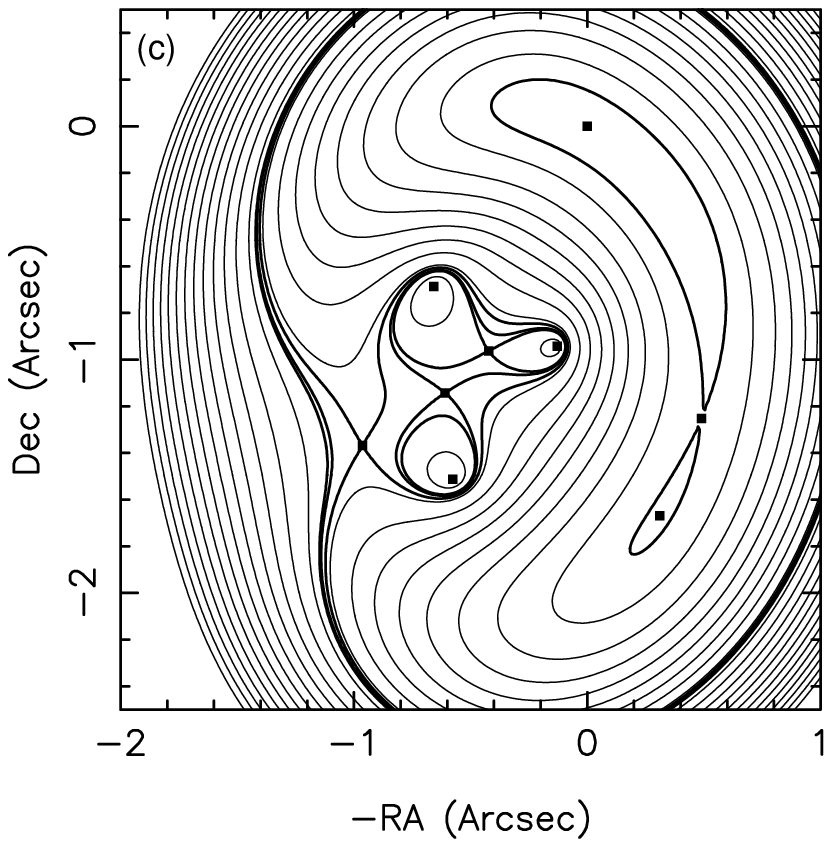,width=3.3in} & \\ 
\end{tabular}
\figurenum{6}
\caption{Best-fit 3SIE model. Plots are the same as in Fig.\ 5. Note that the
critical curve (a) passes closer to A than either B or C, ensuring that A is
the brightest image. The high magnification of A is due to the position of the
unlensed source, which lies just outside a swallowtail cusp (b).}

\end{figure*}

We now turn to the truncated mass distributions. The 3PJS model has a fit
statistic of $\chi^2_{tot} = 538$, a significant improvement over the 3SIS
model but with the same number of free parameters (NDF = 4). The flux density
of A is 11.2 mJy compared to the observed value of 22.6 mJy, but overall the
flux densities are fit rather well ($\chi^2_{flx} = 46$). The fit statistic is
dominated by offsets from the galaxy positions ($\chi^2_{gal} = 290$). The
3PJE/FIX model also does better than its SIE counterpart, offering a fit of
$\chi^2_{tot} = 100$ for NDF = 7. This is dominated by the fit to the flux
density of A (predicted 8.8 mJy, $\chi^2_{flx,A} = 37$). The 3PJE model offers
a best-fit of $\chi^2_{tot} = 65$, but again, A is not the brightest image
(predicted 11.3 mJy, $\chi^2_{flx,A} = 25$). It is interesting to note that
while the 3SIE model more accurately reproduces the flux density of A, the
predicted galaxy position angles in the 3PJE model ($69^{\circ}$,
$18^{\circ}$, $62^{\circ}$ for G, G$'$ and G$''$, respectively) are much
closer to those of the observed surface brightness distributions
($48^{\circ}$, $29^{\circ}$, $63^{\circ}$, respectively). However, the
external shears in the 3PJE and 3PJE/FIX models ($\gamma \simeq 0.25$) are
significantly larger than those in the 3SIE and 3SIE/FIX models ($\gamma
\simeq 0.15$).

To test whether the large external shears predicted by each of our lens models
may be due to a dark matter halo associated with the galaxy group, we also
investigated models with four distinct mass distributions. The external shear
was set to zero in these trials. Although one might expect the three primary
lenses to define the group center, we were unable to find reasonable models in
which the group halo resides within the Einstein ring. Problems arise because
a fourth mass distribution that is even moderately concentrated significantly
disturbs the caustic structure of the system, unless it is centered on one of
the lensing galaxies. This makes it very difficult to account for the image
properties without creating additional lensed components. Not surprisingly, we
find that in the vast majority of our trials the group halo is pushed out of
the Einstein ring toward the east, where it becomes a large external shear
contributor. 

We can check whether the mass ratios of the galaxies, as estimated from their
lens model critical radii $b$, are consistent with the observed luminosity
ratios.  In standard lens models and observations of lenses, $b \propto
L^{1/2}$ (see Keeton, Kochanek \& Falco 1998), so we predict critical radius
ratios of 1.0:$0.62\pm0.2$:$0.63\pm0.2$ for G:G$'$:G$''$ based on the I-band
luminosities in Table 2.  The model ratios -- 1.0:0.59:0.86 (3SIS),
1.0:0.62:0.75 (3SIE/FIX), 1.0:0.80:0.82 (3SIE), 1.0:0.85:0.55 (3PJS),
1.0:0.68:0.60 (3PJE/FIX), and 1.0:0.57:0.68 (3PJE) -- are remarkably
consistent with these predictions. This provides additional evidence that all
three lensing galaxies are important to the model and indicates that the
B-band (for $z_l\simeq 1$) mass-to-light ratios of the three lenses must be
very similar. The compatible galaxy colors and the consistency of our
fundamental plane redshift estimates further reinforce this argument.

\section{Discussion}

HST and VLBA observations demonstrate that the gravitational lens B1359+154
consists of six images of a single background radio source. This is the first
example of a galaxy-scale lens system with more than four images.  The unique
configuration is produced by the complex mass distribution of the lens, a
compact group of three $z_l\simeq 1$ galaxies lying on the vertices of a
triangle separated by $0\farcs7 \simeq 4 h^{-1}$kpc, inside the $1\farcs7$
diameter Einstein ring defined by the radio components. The outer images
(A--D) have the morphology of a standard four-image lens, but the triangle of
galaxies produces additional extrema in the gravitational potential at which
two central images (E and F) form. Simple lens models consisting of three
deflectors constrained by HST galaxy coordinates and VLBA radio component data
naturally produce six images at the observed positions, but require
finely-tuned ellipticities to account for the flux density ratios.  While it
would be premature to claim that we have accurately represented the lensing
mass distribution ($\chi^2_{tot}$/NDF $\geq 10$ for all models), our initial
modeling efforts demonstrate that B1359+154 can be explained as a true
six-image lens system produced by three lensing galaxies. The consistency
between the model-predicted deflector mass ratios and the observed luminosity
ratios of the lens galaxies adds credibility to our results.

The three primary lens galaxies appear to constitute the core of a compact
group. The most concentrated groups with small numbers of galaxies are the
Hickson groups, a quarter of which have 3 members (Hickson et al. 1992).
Separated by only $\simeq 4 h^{-1}$ kpc, the three galaxies are close enough
so that they might be expected to be consumed in a merger on a time scale of
$\sim 0.1$ $H_o^{-1}$ (Barnes 1985). Thus the incidence of high multiplicity
arcsecond-scale lens systems may allow for an estimate of the merger rate of
massive galaxies at $z \simeq 0.5 -1.5$.

The relative simplicity of B1359+154 compared to the lenses found in the cores
of rich clusters should make this system an excellent tool for studying galaxy
halos in dense environments, and the relationship between the galaxies and
their parent halos. One issue worthy of investigation is whether the galaxy
mass distributions are truncated on the scale of inter-galaxy separation. The
improved fits of the 3PJS and 3PJE/FIX models relative to the 3SIS and
3SIE/FIX models offer tentative arguments to this effect. Because predicted
time delays in the PJ models are significantly higher than those in the SIE
models, measured delays may be used to discriminate between the profiles. A
second interesting issue is the placement of the group halo relative to the
lensing galaxies. The image positions and flux densities can be reproduced by
three mass distributions associated with the observed galaxies, so there is
little need to add a fourth independent mass distribution inside the Einstein
ring based on modeling arguments. A fourth deflector above the critical
density for multiple imaging would place an additional radial caustic onto the
source plane and disturb the delicate balance needed to account for the data
without creating additional images.  Thus if this halo exists within the
Einstein ring, it is likely to be either subcritical (large core radius) or
centered on one of the galaxies. However, the similar model-predicted
mass-to-light ratios of the galaxies argue against the latter possibility. The
large external shear fields ($\gamma \gtorder 0.15$) required by all of our
models suggests that a group halo may be displaced from the primary lenses. It
is interesting to note that the orientation of the shear axis is remarkably
constant, running nearly east to west regardless of the deflector model
employed.

The six lensed components allow for more complicated models than we have
investigated here, but the development of a truly robust model of the lensing
mass distribution requires the acquisition of additional constraints. The
detection of lensed extended emission from the quasar host galaxy in both the
HST optical and CFHT infrared images is particularly promising.  Deep,
high-resolution HST imaging using either WFPC2 or NICMOS could investigate the
properties of the Einstein ring in detail and provide vital new constraints on
the potential structure (see Kochanek et al.\ 2000b). Furthermore, if
subcomponents of D, E and F can be detected using deep global VLBI imaging,
the relative orientations of the radio substructure could place additional
constraints on the local magnification matrices. Finally, the measurement of
differential time delays would also offer important constraints on the mass
model, as these directly probe the lensing potential at the image positions.
One might expect the source to be variable, based on the compactness of the
radio components, though there is no evidence for significant variability at
this point. Radio snapshots of B1359+154 should be routinely obtained to
determine whether the variability is large enough to undertake a monitoring
program using current instruments. Otherwise, the enhanced VLA may be able to
determine time delays from the component light curves even if the source
variability is only at the few percent level.

It seems unlikely that the lens is a chance projection of three galaxies (an
{\it a posteriori} probability of $\sim 10^{-3}$ in a sample of 60 lenses),
and their statistically identical colors further argues against this
possibility. While the Next Generation Space telescope could directly
determine the redshifts of all three galaxies, current telescopes can only
measure the average redshift of the lenses unless they have narrow emission
lines that are not apparent in the Myers et al. (1999) spectrum. Two
possibilities are to search for Mg{\sc~II} metal absorption lines in the
source spectrum or HI absorption features in the radio continuum once the
average redshift is known.  For $z_l\simeq1$ the Mg{\sc~II} absorption
features would lie in the V-band where the source flux is 23~V~mag, so the
observation is difficult.  Many groups have significant HI masses (Hickson
1997) and the radio flux density of $\simeq 90$~mJy at 0.7~GHz (based on the
1.7 GHz VLBA flux density and a spectral index of $\alpha \simeq 0.3$ at low
frequency) may be high enough to search for absorption features.  X-ray
observations are less promising because very deep observations would be
required to detect the group (a bright $10^{42}$~erg~s$^{-1}$ group at $z_l=1$
has an X-ray flux of only $4\times 10^{-16}$~ergs~cm$^{-2}$~s$^{-1}$) and the
high resolution of the Chandra Observatory would be needed to distinguish
emission from the group and the lensed sources.

\noindent Acknowledgements: The authors deeply thank the director of STScI,
Steve Beckwith, for making these observations possible.  We also thank Lars
Hernquist and Ann Zabludoff for discussions about groups and the implications
of this lens.  DR acknowledges support from the Zaccheus Daniel
Foundation. Support for the CASTLES project was provided by NASA through grant
numbers GO-8175, GO-8268 and GO-8804 from the Space Telescope Science
Institute, which is operated by the Association of Universities for Research
in Astronomy, Inc.  CSK, EEF, JL, BAM and JAM were also supported by the
Smithsonian Institution.  CSK and CRK were also supported by the NASA
Astrophysics Theory Program grant NAG5-4062.  HWR is also supported by a
Fellowship from the Alfred P. Sloan Foundation.

\appendix

\section{The Statistics of Perturbing Galaxies}

Here we expand on the discussions of the statistics of perturbing galaxies of
gravitational lenses in Keeton, Kochanek \& Seljak (1997) and Kochanek \&
Apostolakis (1988) to include the effects of the galaxy luminosity function.
We also emphasize the important distinction between a primary lens galaxy
possessing a nearby neighbor and that neighbor significantly perturbing either
the lens model or the caustic structure.  We consider only galaxies which are
clustered with the primary lens galaxy. Galaxies and clusters along the line
of sight are an important source of weak shear perturbations but clustered
galaxies dominate the perturbations (see Keeton et al. 1997; Barkana 1996).  A
key distinction to bear in mind is that {\it all } lens galaxies will have
close neighbors (or satellites) if examined closely, so it is only the degree
to which they modify the caustic structure or perturb the images of the lens
which is relevant.  Moreover, the $0.1-1\%$ astrometric precision of most lens
observations permits the detection of very small perturbations to lens models
unless they are degenerate with other properties of the mass model.  The
sensitivity of models to small perturbations means that neighbors may be
important to all lens models while remaining unimportant for most statistical
properties of lens systems (image separations, image multiplicities, cross
sections, etc.).\footnote{ We will discuss only pairs of galaxies.  While the
probability of three galaxies can be determined from the three-point
correlation function, it has never been measured on the kpc scales
corresponding to the separations of galaxies in compact groups.  An empirical
determination based on a local redshift survey would probably be the most
accurate approach to making the estimate.  }

Following Keeton et al.\ (1997) and Kochanek \& Apostolakis (1988) we model
the distribution of neighbors using the correlation function,
$\xi(r)=(r/r_0)^{-\chi}$ with $r_0\simeq 5h^{-1}$~Mpc and $\chi\simeq 1.75$
(e.g.\ Peebles 1993), which when projected onto the lens plane becomes
$\xi_2(R) \simeq 4 r_0^\chi R^{1-\chi}$. We assume that the correlation
function can be extrapolated to the small physical scales relevant to the
calculations.  We will add the effects of the distribution of galaxies in
luminosity by including a luminosity function, $dn/dL=n_*(L/L_*)^\alpha
\exp(-L/L_*)$ where $n_*\simeq 0.01 (h\hbox{Mpc})^{-3}$ and $\alpha\simeq-1$
(e.g.\ Kochanek 1996). We use an SIS lens model to estimate the critical
radius $b= b_* (L/L_*)^{1/2} D_{LS}/D_{OS}$ and the Faber-Jackson relation to
convert between luminosity and velocity dispersion,
$L/L_*=(\sigma/\sigma_*)^4$.  $D_{OL}$, $D_{OS}$ and $D_{LS}$ are standard
angular diameter distances and $b_* = 4\pi (\sigma_*/c)^2 = 1\farcs45$ for
early-type lens galaxies (Kochanek et al.\ 2000a).  The characteristic scale
for finding neighboring galaxies is given by the probability of finding a
galaxy with a comoving density $n_*$ inside the critical radius $b_*$ of an
$L_*$ lens galaxy,
\begin{eqnarray}
   \tau_0 &= &2\pi n_* \int_0^{b_* D_{OL}} R dR \xi_2(R) 
          \simeq 20 n_* r_0^\chi (b_* D_{OL})^{3-\chi}  \\
          &\simeq &0.005 \left[ { n_* \over 0.01 (h\hbox{Mpc})^{-3} }\right] 
               \left[ { r_0 \over 5 h^{-1}\hbox{Mpc} }\right]^{1.75}
               \left[ { b_* D_{OL} \over 5 h^{-1}\hbox{kpc} } \right]^{1.25}.
\end{eqnarray}
The probability of finding a galaxy brighter than luminosity $L$ is
\begin{equation}
   \tau (>L) = \tau_0 \int_{L/L_*}^\infty d\lambda \exp(-\lambda) \lambda^\alpha
\end{equation}
where $\lambda=L/L_*$. The integral logarithmically diverges as $L\rightarrow
0$ for $\alpha=-1$, but the divergence is slow, as the integral is $0.2$,
$1.8$ and $4.0$ for $L/L_*=1$, $0.1$ and $0.01$, respectively.  The result
(A3) is an underestimate for several reasons.  First, early-type galaxies are
more strongly clustered than galaxies in general, so our value of $r_0$ is
low.  Second, for detecting neighbors we should include both early-type and
late-type galaxies, and the total density of all types of galaxies is larger
than $0.01 (h^{-1}\hbox{Mpc})^{-3}$.

The third problem is that the presence of multiple galaxies modifies the cross
sections for finding the system as a gravitational lens.  Kochanek \&
Apostolakis (1988) showed that for galaxies with equal critical radii $b_0$
separated by distance $\ell$, the total cross-section is approximately $2\pi
b_0^2$ for $\ell > 4 b_0$, $\pi b_0^2 \left[2 - (2-l/2b_0)\right]^2$ for $2
b_0 < \ell < 4 b_0$ and $\pi(2b_0-\ell/2)^2$ for $\ell < 2b_0 $.  The
cross-section has a {\it minimum} at $\ell = 2b_0$ where the centers of the
two lenses are deflected to the same point on the source plane.  Where the
total cross-section is smallest, the cross-section for more than two images
and the average image magnification are largest.  In short, as the
cross-section is dropping, the amount of magnification bias will rise in
compensation.  For two galaxies with critical radii $b_1$ and $b_2$ separated
by $\ell$ we can derive several general results.  There are no critical lines
on the axis separating the two galaxies once $\ell < (b_1^{1/2}+b_2^{1/2})^2$,
while the lens can generate images bracketing both lenses once $\ell <
2(b_1+b_2)$.  If the separation is smaller than $\ell <
2\,\hbox{max}(b_1,b_2)$ then it is possible to find solutions with no images
in between the two lenses.  Kochanek et al.\ (2000b) used this to find a model
for the four-image lens B1608+656 using two singular galaxies.  The total
multiple image cross-section is $\pi(b_1^2+b_2^2)$ when the lenses are well
separated, rising to $\pi (b_1+b_2)^2$ when they are perfectly aligned.  The
dependence of the cross section on separation should be similar to the equal
critical radius case, with a minimum total cross-section and a maximum large
image multiplicity cross-section at an intermediate separation $\ell \simeq
(b_1+b_2)$.  Because $b \propto L^{1/2}$, relatively low luminosity galaxies
can produce substantial cross-section enhancements -- a $L_*$ and $0.1L_*$
galaxy have a cross-section when aligned which is 60\% higher than that for
the two galaxies separately.  We can estimate the magnitude of the effect by
computing the fractional ``excess'' cross-section $2\pi b_1b_2/\pi b_1^2$ for
aligned lenses over well separated lenses, assuming that it holds for all
lenses separated by less than $\ell < 2(b_1+b_2)$.  Defining $u=b_1/b_*$ the
fractional excess cross-section from less massive neighboring galaxies ($b_2 <
b_1$) is of order
\begin{equation}
   \simeq 2^{4-\chi}\tau_0 \int_0^{u^2} d\lambda \lambda^{\alpha+1/2} \exp(-\lambda) (u+\lambda^{1/2})^{3-\chi}/u.
\end{equation}
For $u=1$ ($b_1=b_*$) the excess cross-section is $11 \tau_0\sim 6\% $, while
for $u=0.1$ it is only $0.9\tau_0$.  While a numerical calculation including
the effects of magnification bias is needed to obtain the exact impact of
neighbors on lensing statistics, this order of magnitude calculation shows
that the enhancements are modest.
 
Keeton et al. (1997) considered a very simple model for the importance of
neighboring galaxies by estimating the shear that a companion can produce at
the center of the primary lens galaxy.  A neighbor of luminosity $L$ located
at projected radius $R$ produces a shear $\gamma =
(b_*D_{OL}/2R)(L/L_*)^{1/2}$ at the center of the primary lens.  The
probability of finding a shear larger than $\gamma$ is
\begin{equation}
  \tau(>\gamma) \simeq 0.4 \tau_0 \gamma^{-1.25} \int_0^\infty d\lambda
\lambda^{\alpha+(3-\chi)/2} \exp(-\lambda)
\end{equation}   
where $\lambda=L/L_*$. The integral converges (for standard parameters
$\alpha+(3-\chi)/2 \simeq -0.38$ and the value of the integral is $1.4$), and
the perturbations are weakly dominated by low luminosity galaxies.  While the
probability of a lens having a neighbor diverges, the probability that the
neighbor produces a large perturbation on the lens is finite and relatively
small, $\tau(>\gamma) \simeq 0.6\tau_0/\gamma^{1.25}$.  This estimate of the
perturbations makes little sense for perturbing galaxies inside the Einstein
ring or for estimating the perturbations to images at the Einstein radius.  If
we instead estimate the shear produced at the closest point on the unperturbed
critical line of radius $b_0$ of the primary lens, we find
\begin{equation}
  \tau(>\gamma) \simeq \tau_0 \left( {b_0 \over b_*}\right)^{1.25} 
      \int_0^\infty d\lambda \lambda^\alpha \exp(-\lambda) 
      \left| 1 - \left( 1 \pm { b_* \lambda^{1/2} \over 2 \gamma b_0}  \right)^{1.25}\right|
\end{equation}   
where the $+$ ($-$) branch corresponds to perturbations from outside (inside)
the ring.  The integral is not analytic, but under the approximation that
$|1-(1\pm x)^a|=x^a$ we obtain the same result as in A3. The Keeton et
al. (1997) approximation underestimates the total perturbation by about a
factor of 2.  The new approximation still overestimates the contribution from
images inside the Einstein ring, because a perturbing galaxy located at the
center is counted as producing a shear perturbation at the Einstein ring when
its effects are indistinguishable from the monopole of the primary lens.

\clearpage


\begin{deluxetable}{c c c c c c}
\scriptsize
\tablecaption{VLBA Data}
\tablewidth{0pt}
\tablehead{Comp & $\Delta$RA ($''$) & $\Delta$Dec ($''$) & $S_{1.7}$ (mJy) & $b_{maj}$ (mas) &
$\alpha_{1.7}^{5}$}
\startdata
A1 &              $\equiv 0$ &             $\equiv 0$ & 22.6 &  9 & $-0.29$ \nl
B1 & $-0.490398 \pm 0.00003$ & $-1.252428\pm 0.00003$ & 14.5 &  6 & $-0.32$ \nl
C1 & $-0.311420 \pm 0.00002$ & $-1.669560\pm 0.00002$ & 19.8 &  6 & $-0.27$ \nl
D1 & $+0.962772 \pm 0.00010$ & $-1.368612\pm 0.00010$ &  4.5 &  5 & $-0.32$ \nl
E1 & $+0.608880 \pm 0.00008$ & $-1.142871\pm 0.00008$ &  6.0 &  3 & $-0.23$ \nl
F1 & $+0.422273 \pm 0.00025$ & $-0.963759\pm 0.00025$ &  1.8 &  3 & $-0.38$ \nl
A2 & $-0.013783 \pm 0.00013$ & $-0.006793\pm 0.00013$ &  3.3 & 16 & \nl
B2 & $-0.481571 \pm 0.00032$ & $-1.292351\pm 0.00032$ &  1.4 &  8 &\nl
C2 & $-0.340726 \pm 0.00034$ & $-1.622677\pm 0.00034$ &  1.3 &  7 & \nl
\enddata
\tablecomments{Model-fit components for B1359+154 1.7 GHz VLBA data (combined
epochs). Positions are offset from RA $14^h$ $01^m$ $35^s.5476$, Dec $+15'$
$13''$ $25\farcs646$ (J2000). Nominal uncertainties on the flux densities are
three times the map rms of $45$ $\mu$Jy/beam. Positional uncertainties are
taken as the beam size ($\sim 10$ mas) divided by the signal-to-noise for each
component. The major axes of the gaussian components are given by
$b_{maj}$. Spectral indices $\alpha_{1.7}^{5}$ between 1.7 and 5 GHz are
calculated using the combined flux density of the primary and subcomponents,
where applicable. The 5 GHz data is from Rusin et al.\ (2000a). Assuming 5\%
errors on both the 1.7 and 5 GHz flux densities due to variability and
model-fitting, the error on the spectral indices is $\simeq 0.07$.}
\end{deluxetable}

\clearpage


\begin{deluxetable}{lrrccccc}
\scriptsize
\tablecaption{HST Astrometry and Photometry}
\tablewidth{0pt}
\tablehead{ID & $\Delta$RA ($''$) & $\Delta$Dec  ($''$)& I (mag) & V--I (mag) & $R_e$ ($''$) &
$e$ & PA ($^\circ$)}
\startdata
 A        &   $\equiv$0         &   $\equiv$0           &$24.01\pm0.06$   &$0.68\pm0.12$   &   &&  \nl
 B        & --0.483$\pm$0.007   &   --1.253$\pm$0.009   &$24.29\pm0.11$   &$0.43\pm0.15$   &   &&  \nl
 C        & --0.323$\pm$0.007   &   --1.640$\pm$0.003   &$24.01\pm0.05$   &$0.27\pm0.13$   &   &&  \nl
 D        &   0.957$\pm$0.008   &   --1.357$\pm$0.008   &$25.19\pm0.08$   &$0.61\pm0.17$   &   &&  \nl
 E        &   0.627$\pm$0.013   &   --1.129$\pm$0.011   &$25.12\pm0.17$   &$0.45\pm0.19$   &   &&  \nl
 F        &   0.426$\pm$0.017   &   --0.951$\pm$0.028   &$26.25\pm0.23$   &$0.29\pm0.23$   &   &&  \nl
 G        &   0.160$\pm$0.008   &   --0.935$\pm$0.024   &$22.68\pm0.28$   &$3.31\pm0.55$   &0.71$\pm$0.29   &0.11$\pm$0.23 &48$\pm$55 \nl
 G$'$     &   0.658$\pm$0.005   &   --0.690$\pm$0.008   &$23.69\pm0.24$   &$3.13\pm0.29$   &0.16$\pm$0.06   &0.69$\pm$0.14 &29$\pm$8  \nl
 G$'$$'$  &   0.575$\pm$0.007   &   --1.516$\pm$0.011   &$23.70\pm0.33$   &$3.23\pm1.01$   &0.14$\pm$0.07   &0.49$\pm$0.20 &63$\pm$7   \nl
\enddata
\tablecomments{The three lensing galaxies are fit to de Vaucouleurs
profiles. The effective radii ($R_e$), ellipticity parameters $e = 1 - f$
and position angles (PA) of these fits are listed.}
\end{deluxetable}

\begin{deluxetable}{lrrccccrl}
\footnotesize
\tablecaption{Nearby Objects}
\tablehead{ID  & $\Delta$RA ($''$) & $\Delta$Dec ($''$) & I & V--I & I--K & $\gamma$ & $\theta_\gamma$ & Notes}
\startdata
 G     &   0.2 &  --0.9 & $22.68\pm0.55$ &  $3.31\pm0.97$ & $\sim$2.0 &        &      & K3, Group\\
 G$'$  &   0.7 &  --0.7 & $23.69\pm0.14$ &  $3.13\pm0.35$ & $\sim$3.0 &        &      & K1, Group\\
 G$''$ &   0.6 &  --1.5 & $23.70\pm0.12$ &  $3.23\pm1.03$ & $\sim$3.0 &        &      & K2, Group\\
 S1   &    4.4 &    9.3 & $24.39\pm0.11$ &     $>1.74$    &           &        &      & \\
 S2   &   11.5 &  --2.9 & $24.94\pm0.12$ &     $>1.43$    &           &        &      & \\
 G1   & --13.5 &  --7.5 & $19.08\pm0.10$ &  $1.36\pm0.03$ &           &  0.076 &   64 & \\
 G2   & --16.8 &    1.7 & $22.73\pm0.10$ &  $1.42\pm0.05$ &           &  0.013 & --81 & \\
 G3   &    0.6 & --10.6 & $23.30\pm0.10$ &  $0.79\pm0.04$ &           &  0.017 &  --3 & \\
 G4   &    2.1 &    5.2 & $23.52\pm0.11$ &  $0.94\pm0.06$ &           &  0.023 &   18 & \\
 G5   &    3.7 &  --1.3 & $23.52\pm0.11$ &     $>2.41$    & $\sim$4.4 &  0.042 & --85 & K4, Group\\
 G6   & --17.5 &    6.7 & $23.58\pm0.11$ &  $1.09\pm0.06$ &           &  0.008 & --67 & \\
 G7   &  --9.1 &  --5.1 & $23.97\pm0.11$ &  $1.17\pm0.06$ &           &  0.012 &   66 & \\
 G8   &  --7.1 &    8.4 & $24.31\pm0.11$ &  $1.48\pm0.08$ &           &  0.009 & --38 & \\
 G9   &  --1.8 &    1.5 & $24.50\pm0.12$ &     $>1.37$    & $\sim$3.5 &  0.030 & --38 & K5, Group\\
 G10  &    6.8 & --12.2 & $24.70\pm0.12$ &  $0.63\pm0.09$ &           &  0.007 & --31 & \\
 G11  &   12.2 &  --2.6 & $24.80\pm0.12$ &  $1.54\pm0.09$ &           &  0.007 & --82 & \\
 K6   &  --6.5 &  --6.9 & $>25.0$        &                & $>$4      &        &      & K6\\
 T1   & --32.3 & --67.8 & $18.26\pm0.10$ &  $1.38\pm0.03$ &           &  0.023 &   26 & \\
 T2   & --47.0 &  --9.9 & $18.85\pm0.10$ &  $2.20\pm0.03$ &           &  0.027 &   79 & \\
 T3   & --34.0 & --50.1 & $19.23\pm0.10$ &  $2.12\pm0.03$ &           &  0.018 &   35 & \\

\enddata \tablecomments{SExtractor position offsets from A are given in
arcseconds. G,G$'$,G$''$ photometry and positions are from component fits in
Table~2. Galaxies on the WFPC2 image within 20\arcsec\ of G are labeled G$x$,
and those detected only in the Rusin et al.\ (2000a) infrared observations are
labeled K$x$.  Stars are labeled as S$x$. Galaxies outside of 20\arcsec\ with
large estimated shears are included as T$x$.  Tidal shear estimates $\gamma$
assume SIS halos with the same redshift and mass-to-light ratio as G, and the
shear position angles $\theta_\gamma$ are relative to G in degrees
counter-clockwise from North (see Leh\'ar et al. 2000).  }
\end{deluxetable}

\def\mc#1{\multicolumn{1}{c}{#1}}
\begin{deluxetable}{llccccrcrl}
\footnotesize \tablecaption{Lens Models} \tablehead{Type &Lens &$b$ &$x_l$
&$y_l$ &$e,\gamma$ &$PA$ &Img &\mc{$\Delta t_{iA}$} &$\chi^2$ \\ &
&$('')$ &$('')$ &$('')$ & &($^\circ$) & &\mc{$(h^{-1}\hbox{days})$} & }
\startdata \input modtab \enddata
\tablecomments{The five lines for each model give the properties of the
lensing galaxies (G, G$'$ and G$''$), external shear, and recovered source
coordinates: $b$ is the critical radius,
$x_l$ and $y_l$ are the lens coordinates, $e=1-f$ where $f$ is the axis ratio
of the ellipsoid, $\gamma$ is the strength of the external shear, and PA is
the major axis position angle of the ellipsoid or the direction (to a tidal
perturbation) producing the shear.  The time delays for each image are defined
relative to image A as $\Delta t_{iA}=t_i-t_A$ and were computed for an
$\Omega_0=0.3$ flat universe and a lens redshift $z_l=1$. The final column
gives $\chi^2_{tot}$ and the contributions from the astrometry of the images
($\chi^2_{pos}$), the flux ratios of the images ($\chi^2_{flx}$) and the
positions of the lens galaxies ($\chi^2_{gal}$). The models assume that the
uncertainties in the positions of the radio sources are 1 mas and that the
fractional uncertainties in their flux densities are 10\%.}
\end{deluxetable}







\end{document}

%% file: modtab.tex
3SIS     &G       &$ 0.400$ &$ 0.179$ &$-0.906$ &$ 0.000$ &$   0.0$ &B &$ 9.7$ &$\chi^2_{tot}= 1569.8$\\
         &G$'$    &$ 0.236$ &$ 0.680$ &$-0.734$ &$ 0.000$ &$   0.0$ &C &$ 9.3$ &$\qquad\chi^2_{pnt}= 1341.6$\\
         &G$''$   &$ 0.344$ &$ 0.567$ &$-1.559$ &$ 0.000$ &$   0.0$ &D &$35.8$ &$\qquad\chi^2_{flx}= 154.6$\\
         &shear   &$      $ &$      $ &$      $ &$ 0.154$ &$  87.9$ &E &$36.8$ &$\qquad\chi^2_{gal}=  73.6$\\
  	 &src     &$      $ &$ 0.330$ &$-1.018$ &         &         &F &$36.9$ & \\
\hline

3SIE/FIX &G       &$ 0.409$ &$ 0.135$ &$-0.912$ &$ 0.170$ &$  48.0$ &B &$ 9.6$ &$\chi^2_{tot}= 215.8$\\
         &G$'$    &$ 0.253$ &$ 0.670$ &$-0.718$ &$ 0.000$ &$  29.0$ &C &$ 9.2$ &$\qquad\chi^2_{pnt}=  3.9$\\
         &G$''$   &$ 0.306$ &$ 0.581$ &$-1.506$ &$ 0.139$ &$  63.0$ &D &$33.0$ &$\qquad\chi^2_{flx}=  181.6$\\
         &shear   &$      $ &$      $ &$      $ &$ 0.123$ &$  97.4$ &E &$34.0$ &$\qquad\chi^2_{gal}=  30.3$\\
  	 &src     &$      $ &$ 0.328$ &$-0.991$ &         &         &F &$34.3$ & \\
\hline

3SIE     &G       &$ 0.359$ &$ 0.128$ &$-0.944$ &$ 0.338$ &$ -10.6$ &B &$ 1.0$ &$\chi^2_{tot}= 43.2$\\
         &G$'$    &$ 0.287$ &$ 0.658$ &$-0.688$ &$ 0.589$ &$ -47.6$ &C &$ 0.8$ &$\qquad\chi^2_{pnt}=  0.2$\\
         &G$''$   &$ 0.294$ &$ 0.576$ &$-1.513$ &$ 0.416$ &$ -78.1$ &D &$24.7$ &$\qquad\chi^2_{flx}=  27.0$\\
         &shear   &$      $ &$      $ &$      $ &$ 0.132$ &$  71.5$ &E &$25.3$ &$\qquad\chi^2_{gal}=  16.0$\\
  	 &src     &$      $ &$ 0.352$ &$-0.980$ &         &         &F &$25.5$ & \\
\hline

3PJS     &G       &$ 1.371$ &$ 0.127$ &$-1.030$ &$ 0.000$ &$   0.0$ &B &$17.6$ &$\chi^2_{tot}= 537.9$\\
         &G$'$    &$ 1.170$ &$ 0.618$ &$-0.693$ &$ 0.000$ &$   0.0$ &C &$16.6$ &$\qquad\chi^2_{pnt}= 202.1$\\
         &G$''$   &$ 0.754$ &$ 0.649$ &$-1.416$ &$ 0.000$ &$   0.0$ &D &$66.5$ &$\qquad\chi^2_{flx}=  45.8$\\
         &shear   &$      $ &$      $ &$      $ &$ 0.235$ &$  96.2$ &E &$69.3$ &$\qquad\chi^2_{gal}= 290.0$\\
  	 &src     &$      $ &$ 0.223$ &$-0.919$ &         &         &F &$71.2$ & \\
\hline

3PJE/FIX &G       &$ 1.348$ &$ 0.109$ &$-0.964$ &$ 0.171$ &$  48.0$ &B &$20.5$ &$\chi^2_{tot}= 99.8$\\
         &G$'$    &$ 0.914$ &$ 0.664$ &$-0.700$ &$ 0.448$ &$  29.0$ &C &$19.7$ &$\qquad\chi^2_{pnt}=  3.9$\\
         &G$''$   &$ 0.812$ &$ 0.581$ &$-1.488$ &$ 0.506$ &$  63.0$ &D &$66.1$ &$\qquad\chi^2_{flx}=  44.0$\\
         &shear   &$      $ &$      $ &$      $ &$ 0.255$ &$  94.0$ &E &$66.7$ &$\qquad\chi^2_{gal}=  51.9$\\
  	 &src     &$      $ &$ 0.242$ &$-0.941$ &         &         &F &$68.2$ & \\
\hline

3PJE     &G       &$ 1.095$ &$ 0.134$ &$-0.962$ &$ 0.581$ &$  69.2$ &B &$27.5$ &$\chi^2_{tot}= 64.7$\\
         &G$'$    &$ 0.626$ &$ 0.658$ &$-0.693$ &$ 0.665$ &$  18.4$ &C &$25.8$ &$\qquad\chi^2_{pnt}=  1.1$\\
         &G$''$   &$ 0.745$ &$ 0.580$ &$-1.489$ &$ 0.603$ &$  62.0$ &D &$73.9$ &$\qquad\chi^2_{flx}=  45.1$\\
         &shear   &$      $ &$      $ &$      $ &$ 0.262$ &$  95.8$ &E &$73.6$ &$\qquad\chi^2_{gal}=  18.5$\\
  	 &src     &$      $ &$ 0.230$ &$-0.929$ &         &         &F &$74.9$ & \\
\hline